%% file: template.tex
\documentclass[journal,article,accept,preprints,pdftex,moreauthors]{Definitions/mdpi} 

\firstpage{1} 
\pubvolume{1}
\issuenum{1}
\articlenumber{0}
\pubyear{2022}
\copyrightyear{2022}
\datereceived{} 
\dateaccepted{} 
\datepublished{} 
\hreflink{https://doi.org/} 

\usepackage{amssymb}
\usepackage{amsmath}
\usepackage{balance}
\usepackage{gensymb}
\usepackage[at]{easylist}
\usepackage{xcolor, colortbl}
\usepackage{graphicx}
\usepackage[normalem]{ulem}
\useunder{\uline}{\ul}{}
\usepackage{caption}
\usepackage{subcaption}
\usepackage{booktabs}

\usepackage{nomencl}
\makenomenclature

\input{acronyms_nomencl}

\usepackage{acro}
\input{acronyms.tex}

\Title{An Analysis of RF Transfer Learning Behavior \\Using Synthetic Data}

\TitleCitation{An Analysis of RF Transfer Learning Behavior Using Synthetic Data}

\Author{Lauren J. Wong $^{1,2,3*}$\orcidA{}, Sean McPherson $^{3}$\orcidB{}, and Alan J. Michaels $^{1,2}$\orcidC{}}

\AuthorNames{Lauren J. Wong, Sean McPherson, and Alan J. Michaels}

\AuthorCitation{Wong L.J.; McPherson S; Michaels A.J.}

\address{%
$^{1}$ \quad Hume Center for National Security and Technology, Virginia Tech\\
$^{2}$ \quad Bradley Department of Electrical and Computer Engineering, Virginia Tech\\
$^{3}$ \quad Intel AI Lab, Santa Clara, CA}

\corres{Correspondence: ljwong@vt.edu}

\abstract{\Ac{TL} techniques, which leverage prior knowledge gained from data with different distributions to achieve higher performance and reduced training time, are often used in \ac{CV} and \ac{NLP}, but have yet to be fully utilized in the field of \ac{RFML}.
This work systematically evaluates how \ac{RF} \ac{TL} behavior by examining how the training domain and task, characterized by the transmitter/receiver hardware and channel environment, impact \ac{RF} \ac{TL} performance for an example \ac{AMC} use-case.
Through exhaustive experimentation using carefully curated synthetic datasets with varying signal types, \acp{SNR}, and \acp{FO}, generalized conclusions are drawn regarding how best to use \ac{RF} \ac{TL} techniques for domain adaptation and sequential learning. 
Consistent with trends identified in other modalities, results show that \ac{RF} \ac{TL} performance is highly dependent on the similarity between the source and target domains/tasks.
Results also discuss the impacts of channel environment, hardware variations, and domain/task difficulty on \ac{RF} \ac{TL} performance, and compare \ac{RF} \ac{TL} performance using head re-training and model fine-tuning methods.}


\begin{document}


\section{Introduction}
\input{text/intro}

\section{Background}\label{sec:related}
\input{text/background}

\input{tables/_signals}

\section{Methodology}\label{sec:methodology}
\input{text/methodology}

\section{Experimental Results \& Analysis}\label{sec:results}
\input{text/results}

\section{Conclusion}\label{sec:conclusion}
\input{text/conclusion}

\vspace{6pt} 



\authorcontributions{Conceptualization, Lauren J. Wong; methodology, Lauren J. Wong, Sean McPherson, and Alan J. Michaels; software, Lauren J. Wong; validation, Lauren J. Wong, Sean McPherson, and Alan J. Michaels; formal analysis, Lauren J. Wong; investigation, Lauren J. Wong; resources, Sean McPherson; data curation, Lauren J. Wong; writing---original draft preparation, Lauren J. Wong; writing---review and editing, Sean McPherson and Alan J. Michaels; visualization, Lauren J. Wong; supervision, Sean McPherson and Alan J. Michaels; project administration, Sean McPherson and Alan J. Michaels; funding acquisition, Not Applicable. All authors have read and agreed to the published version of the manuscript.}

\funding{This research received no external funding.}

\institutionalreview{Not applicable.}

\informedconsent{Not applicable.}

\dataavailability{The dataset used in this work is publicly available on IEEE DataPort at \url{https://ieee-dataport.org/open-access/transfer-learning-rf-domain-adaptation-\%E2\%80\%93-synthetic-dataset}.} 


\conflictsofinterest{The authors declare no conflict of interest.} 



\abbreviations{Abbreviations}{
The following abbreviations are used in this manuscript:\\

\printnomenclature
}




\begin{adjustwidth}{-\extralength}{0cm}

\reftitle{References}
\bibliography{bibliography}

\end{adjustwidth}
\end{document}

%% file: acronyms_nomencl.tex
\nomenclature{AMC}{automatic modulation classification}
\nomenclature{AWGN}{additive white Gaussian noise}
\nomenclature{CLDNN}{convolutional long-short term deep neural network}
\nomenclature{CNN}{convolutional neural network}
\nomenclature{CV}{computer vision}
\nomenclature{DARPA}{Defense Advanced Research Projects Agency}
\nomenclature{DL}{deep learning}
\nomenclature{DNN}{deep neural network}
\nomenclature{IQ}{in-phase/quadrature}
\nomenclature{FO}{frequency offset}
\nomenclature{LEEP}{Log Expected Empirical Prediction}
\nomenclature{LogME}{Logarithm of Maximum Evidence}
\nomenclature{LSTM}{Long Short-Term Memory}
\nomenclature{ML}{machine learning}
\nomenclature{MLP}{multi-layer perceptrons}
\nomenclature{NLP}{natural language processing}
\nomenclature{NN}{neural network}
\nomenclature{RF}{radio frequency}
\nomenclature{RFML}{radio frequency machine learning}
\nomenclature{RNN}{Recurrent Neural Network}
\nomenclature{RRC}{root-raised cosine}
\nomenclature{SEI}{specific emitter identification}
\nomenclature{SNR}{signal-to-noise ratio}
\nomenclature{TL}{transfer learning}

\nomenclature{BPSK}{binary phase-shift keying}
\nomenclature{QPSK}{quadrature phase-shift keying}
\nomenclature{PSK8}{phase-shift keying, order 8}
\nomenclature{PSK16}{phase-shift keying, order 16}
\nomenclature{OQPSK}{offset quadrature phase-shift keying}
\nomenclature{QAM16}{quadrature amplitude modulation, order 16}
\nomenclature{QAM32}{quadrature amplitude modulation, order 32}
\nomenclature{QAM64}{quadrature amplitude modulation, order 64}
\nomenclature{APSK16}{amplitude and phase-shift keying, order 16}
\nomenclature{APSK32}{amplitude and phase-shift keying, order 32}
\nomenclature{FSK5k}{frequency-shift keying, 5kHz carrier spacing}
\nomenclature{FSK75k}{frequency-shift keying, 75kHz carrier spacing}
\nomenclature{GFSK5k}{Gaussian frequency-shift keying, 5kHz carrier spacing}
\nomenclature{GFSK75k}{Gaussian frequency-shift keying, 75kHz carrier spacing}
\nomenclature{MSK}{minimum-shift keying}
\nomenclature{GMSK}{Gaussian minimum-shift keying}
\nomenclature{FM-NB}{narrow band frequency modulation}
\nomenclature{FM-WB}{wide band frequency modulation}
\nomenclature{AM-DSB}{amplitude modulation, double-sideband}
\nomenclature{AM-DSBSC}{amplitude modulation, double-sideband suppressed-carrier}
\nomenclature{AM-LSB}{amplitude modulation, lower-sideband}
\nomenclature{AM-USB}{amplitude modulation, upper-sideband}

%% file: acronyms.tex
\DeclareAcronym{AMC}{
  short = AMC ,
  long  = automatic modulation classification ,
  tag = abbrev
}

\DeclareAcronym{NN}{
  short = NN ,
  long  = neural network ,
  tag = abbrev
}

\DeclareAcronym{FO}{
  short = FO ,
  long  = frequency offset ,
  tag = abbrev
}

\DeclareAcronym{MLP}{
  short = MLP ,
  long  = multi-layer perceptron ,
  tag = abbrev
}

\DeclareAcronym{CNN}{
  short = CNN ,
  long  = convolutional neural network ,
  tag = abbrev
}

\DeclareAcronym{CLDNN}{
  short = CLDNN ,
  long  = convolutional long-short term deep neural network  ,
  tag = abbrev
}

\DeclareAcronym{DNN}{
  short = DNN ,
  long  = deep neural network  ,
  tag = abbrev
}

\DeclareAcronym{RNN}{
  short = RNN ,
  long  = recurrent neural network ,
  tag = abbrev
}

\DeclareAcronym{RRC}{
  short = RRC ,
  long  = root-raised cosine ,
  tag = abbrev
}

\DeclareAcronym{LSTM}{
  short = LSTM ,
  long  = long short-term memory ,
  tag = abbrev
}

\DeclareAcronym{UAV}{
  short = UAV ,
  long  = unmanned aerial vehicle ,
  tag = abbrev
}

\DeclareAcronym{CV}{
  short = CV ,
  long  = computer vision ,
  tag = abbrev
}

\DeclareAcronym{ML}{
  short = ML ,
  long  = machine learning ,
  tag = abbrev
}

\DeclareAcronym{TL}{
  short = TL ,
  long  = transfer learning ,
  tag = abbrev
}

\DeclareAcronym{AE}{
  short = AE ,
  long  = autoencoder ,
  tag = abbrev
}

\DeclareAcronym{RF}{
  short = RF ,
  long  = radio frequency ,
  tag = abbrev
}

\DeclareAcronym{RL}{
  short = RL ,
  long  = reinforcement learning ,
  tag = abbrev
}

\DeclareAcronym{DL}{
  short = DL ,
  long  = deep learning ,
  tag = abbrev
}

\DeclareAcronym{NLP}{
  short = NLP ,
  long  = natural language processing ,
  tag = abbrev
}

\DeclareAcronym{SVM}{
  short = SVM ,
  long  = support vector machine ,
  tag = abbrev
}

\DeclareAcronym{SEI}{
    short = SEI ,
    long = specific emitter identification ,
    tag = abbrev
}

\DeclareAcronym{IQ}{
  short = IQ,
  long  = In-phase/Quadrature ,
  tag = abbrev
}

\DeclareAcronym{SNR}{
    short = SNR ,
    long = signal-to-noise ratio ,
    tag = abbrev
}

\DeclareAcronym{RFML}{
  short = RFML ,
  long  = radio frequency machine learning ,
  tag = abbrev
}

\DeclareAcronym{AWGN}{
  short = AWGN ,
  long  = additive white Gaussian noise ,
  tag = abbrev
}

\DeclareAcronym{DARPA}{
  short = DARPA ,
  long  = Defense Advanced Research Projects Agency ,
  tag = abbrev
}

\DeclareAcronym{IARPA}{
  short = IARPA ,
  long  = Intelligence Advanced Research Projects Activity ,
  tag = abbrev
}

\DeclareAcronym{LEEP}{
  short = LEEP ,
  long  = Log Expected Empirical Prediction ,
  tag = abbrev
}

\DeclareAcronym{LogME}{
  short = LogME ,
  long  = Logarithm of Maximum Evidence ,
  tag = abbrev
}

\DeclareAcronym{RFMLS}{
  short = RFMLS ,
  long  = Radio Frequency Machine Learning Systems ,
  tag = abbrev
}

\DeclareAcronym{SCISRS}{
  short = SCISRS ,
  long  = Securing Compartmented Information with Smart Radio Systems,
  tag = abbrev
}

\DeclareAcronym{DySPAN}{
  short = DySPAN,
  long  = IEEE International Symposium on Dynamic Spectrum Access Networks ,
  tag = abbrev
}

\DeclareAcronym{GBC}{
  short = GBC,
  long  = Gaussian Bhattacharyya Coefficient ,
  tag = abbrev
}

\DeclareAcronym{NCE}{
  short = NCE,
  long  = Negative Conditional Entropy ,
  tag = abbrev
}

\DeclareAcronym{JC-NCE}{
  short = JC-NCE,
  long  = Joint Correspondences Negative Conditional Entropy ,
  tag = abbrev
}

\DeclareAcronym{OTCE}{
  short = OTCE,
  long  = Optimal Transport-based Conditional Entropy ,
  tag = abbrev
}

%% file: text/intro.tex
\Acf{RFML} is loosely defined as the application of \ac{DL} to raw \ac{RF} data, and has yielded state-of-the-art algorithms for spectrum awareness, cognitive radio, and networking tasks.
Existing \ac{RFML} works have delivered increased performance and flexibility, and reduced the need for pre-processing and expert-defined feature extraction techniques.
As a result, \ac{RFML} is expected to enable greater efficiency, lower latency, and better spectrum efficiency in 6G systems \cite{5g}.
However, to date, little research has considered and evaluated the performance of these algorithms in the presence of changing hardware platforms and channel environments, adversarial contexts, or resource constraints that are likely to be encountered in real-world systems \cite{wong2021ecosystem}.

\begin{figure}[t]
    \begin{adjustwidth}{-\extralength}{0cm}
    \centering
    \begin{subfigure}[b]{.55\textwidth}
      \centering
      \includegraphics[width=7 cm]{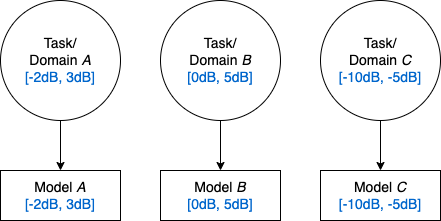}
      \vspace{0.25cm}
      \caption{\centering{Traditional Machine Learning}}
      \label{fig:tl_stock1}
    \end{subfigure}
    \begin{subfigure}[b]{.55\textwidth}
      \centering
      \includegraphics[width=7 cm]{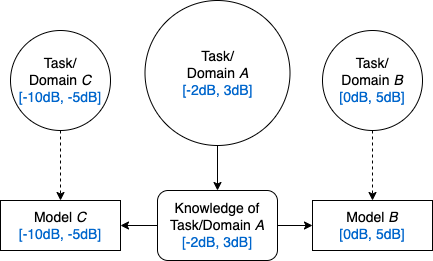}
      \caption{\centering{Transfer Learning}}
      \label{fig:tl_stock2}
    \end{subfigure}
    \caption{In traditional \ac{ML} (Fig. \ref{fig:tl_stock1}), a new model is trained from random initialization for each domain/task pairing. \ac{TL} (Fig. \ref{fig:tl_stock2}) utilizes prior knowledge learned on one domain/task, in the form of a pre-trained model, to improve performance on a second domain and/or task. A concrete example for environmental adaptation to \ac{SNR} is given in blue.}
    \label{fig:tl_stock}
    \end{adjustwidth}
\end{figure}

Current state-of-the-art \ac{RFML} techniques rely upon supervised learning techniques trained from random initialization, and thereby assume the availability of a large corpus of labeled training data (synthetic, captured, or augmented \cite{clark2020}), which is representative of the anticipated deployed environment.
Over time, this assumption inevitably breaks down as a result of changing hardware and channel conditions, and as a consequence, performance degrades significantly \cite{hauser2018, sankhe2019}.
\Ac{TL} techniques can be used to mitigate these performance degradations by using prior knowledge obtained from a \textit{source} domain and task, in the form of learned representations, to improve performance on a ``similar" \textit{target} domain and task using less data, as depicted in Fig. \ref{fig:tl_stock}.

\begin{figure}[t]
    \begin{adjustwidth}{-\extralength}{0cm}
    \centering
    \includegraphics[width=17cm]{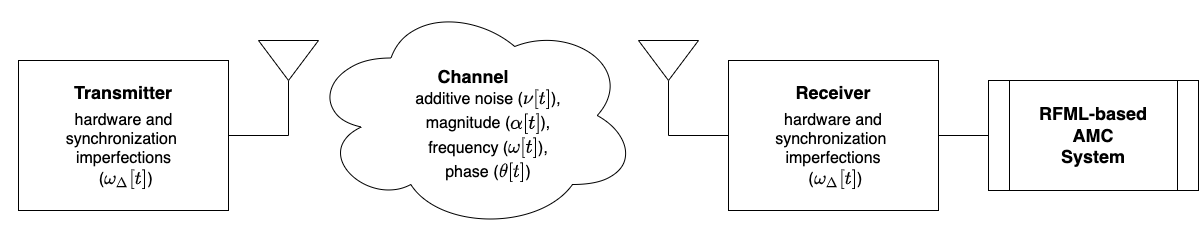}
    \caption{A system overview of the \ac{RF} hardware and channel environment simulated in this work with the parameters/variables ($\alpha[t]$, $\omega[t]$, $\theta[t]$, $\nu[t]$, $\omega_\Delta[t]$) that each component of the system has the most significant impact on.}
    \label{fig:system}
    \end{adjustwidth}
\end{figure}

Though \ac{TL} techniques have demonstrated significant benefits in fields such as \ac{CV} and \ac{NLP} \cite{ruder2019}, including higher performing models, significantly less training time, and far fewer training samples \cite{olivas2009}, \cite{wong2022} showed that the use of \ac{TL} in \ac{RFML} is currently lacking through the construction of an \ac{RFML} specific \ac{TL} taxonomy.
This work begins to address current limitations in understanding how the training domain and task impact learned behavior and therefore facilitate or prevent successful transfer, where the training domain is characterized by the \ac{RF} hardware and the channel environment \cite{wong2022} depicted in Fig. \ref{fig:system} and the training task is the application being addressed including the range of possible outputs (i.e. the modulation schemes classified).
More specifically, this work systematically evaluates \ac{RF} \ac{TL} performance, as measured by post-transfer top-1 accuracy, as a function of several parameters of interest for an \ac{AMC} use-case \cite{hauser2018} using synthetic datasets.
First, \ac{RF} domain adaptation performance is examined as a function of
\begin{easylist}[itemize]
@ \Ac{SNR}, which represents an \textit{environment adaptation} problem characterized by a change in the \ac{RF} channel environment (i.e., an increase/decrease in the additive interference, $\nu[t]$, of the channel) and/or transmitting devices (i.e., an increases/decrease in the magnitude, $\alpha[t]$, of the transmitted signal), 
@ \Ac{FO}, which represents a \textit{platform adaptation} problem characterized by a change in the transmitting and/or receiving devices (i.e., an increase/decrease in $\omega_\Delta[t]$ due to hardware imperfections or a lack of synchronization), and 
@ Both \ac{SNR} and \ac{FO}, representing an \textit{environment platform co-adaptation} problem charactereized by a change in both the \ac{RF} channel environment \emph{and} the transmitting/receiving devices.
\end{easylist}
\noindent Parameter sweeps over these three scenarios addresses each type of \ac{RF} domain adaptation discussed in the \ac{RFML} \ac{TL} taxonomy \cite{wong2022}, and resulted in the construction of 81 training sets, 81 validation sets, and 81 test sets and the training and evaluation of 4360 models.
Additionally, \ac{RF} sequential learning performance is evaluated across broad categories of modulation types, namely linear, frequency-shifted, and analog modulation schemes, as well as in a successive model refinement scenario, where a single modulation type is added/removed from the source dataset.
These experiments resulted in an additional 17 training sets, 17 validation sets, and 17 test sets, and the training and evaluation of 304 models.
From these experiments, we identify a number of practical takeaways for how best to utilize \ac{TL} in an \ac{RFML} setting including how changes in \ac{SNR} and \ac{FO} impact the difficulty of \ac{AMC} and a comparison of head re-training versus fine-tuning for \ac{RF} \ac{TL}.
These takeaways serve as initial guidelines for \ac{RF} \ac{TL}, subject to further experimentation using additional signal types, channel models, use-cases, model architectures, and augmented or captured datasets.

This paper is organized as follows:
Section \ref{sec:related} provides requisite background knowledge of \ac{TL} and \ac{RFML}.
In Section \ref{sec:methodology}, each of the key methods and systems used and developed for this work are described in detail, including the simulation environment and dataset creation, as well as the model architecture and training.
Section \ref{sec:results} presents experimental results and analysis, addressing the key research questions described above.
Finally, Section \ref{sec:conclusion} offers conclusions about the effectiveness of \ac{TL} for \ac{RFML} and next steps for incorporating and extending \ac{TL} techniques in \ac{RFML}-based research.
A list of the acronyms used in this work is provided in the appendix for reference.

%% file: text/background.tex
The following subsections provide an overview of \ac{RFML}, \ac{TL}, and \ac{TL} for \ac{RFML} to provide context for the work performed herein.

\subsection{Radio Frequency Machine Learning (RFML)}
The term \ac{RFML} is often used in the literature to describe any application of \ac{ML} or \ac{DL} to the \ac{RF} domain.
However, \ac{RFML} was coined by \ac{DARPA} and defined as systems that: \begin{easylist}[itemize]
@ Autonomously learn features from raw data to detect, characterize, and identify signals-of-interest, 
@ Can autonomously configure \ac{RF} sensors or communications platforms for changing communications environments, and
@ Can synthesize ``any possible waveform" \cite{rfmls}.
\end{easylist}
\noindent Therefore, \ac{RFML} algorithms typically utilize raw \ac{RF} data as input to \ac{ML}/\ac{DL} techniques; most often \acp{DNN}. 

To date, most \ac{RFML} research has focused on delivering state-of-the-art performance on spectrum awareness and cognitive radio tasks, whether through increased accuracy, increased adaptability, or using less expert knowledge.
Such spectrum awareness cognitive radio tasks include signal detection, signal classification or \ac{AMC}, \ac{SEI}, channel modeling/emulation, positioning/localization, and spectrum anomaly detection \cite{wong2021ecosystem}.
One of the most common and arguably the most mature spectrum awareness or cognitive radio applications explored in the literature is \ac{AMC}, and as such, \ac{AMC} is the example use-case in this work.
\ac{AMC} is the task of identifying the type of or format of a detected signal, and is a key step in receiving \ac{RF} signals.
Traditional \ac{AMC} techinques have typically consisted of an expert-defined feature extraction stage and a pattern recognition stage using techniques such as decision trees, support vector machines, and \acp{MLP} \cite{dobre2007}.
\ac{RFML}-based approaches aim to both automatically learn and identify key features within signals-of-interest, as well as utilize those features to classify the signal, using only minimally pre-processed raw \ac{RF} as input to \ac{DNN} architectures including \acp{CNN} and \acp{RNN} \cite{west2017}.

\begin{figure}[t]
    \begin{adjustwidth}{-\extralength}{0cm}
    \centering
    \includegraphics[width=13.5cm]{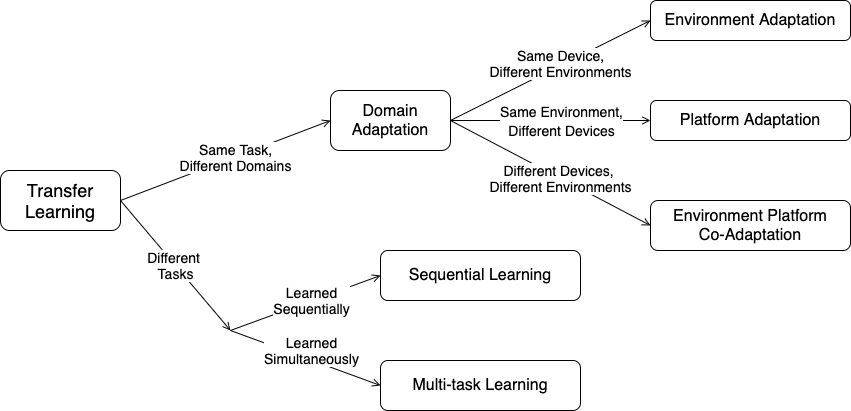}
    \caption{The \ac{RF} \ac{TL} taxonomy proposed in \cite{wong2022}.}
    \label{fig:taxonomy}
    \end{adjustwidth}
\end{figure} 

\subsection{Transfer Learning (TL) for \ac{RFML}}

As previously mentioned, \ac{TL} aims to utilize prior knowledge gained from a source domain/task to improve performance on a ``similar" target domain/task, where training data may be limited.
The domain, $D = \{X, P(X)\}$, consists of the input data $X$ and the marginal probability distribution over the data $P(X)$.
Meanwhile, the task, $T = \{Y, P(Y \vert X)\}$, consists of the label space $Y$, and the conditional probability distribution $P(Y \vert X)$ learned from the training data pairs $\{x_i, y_i\}$ such that  $x_i \in X$ and $y_i \in Y$.
In the context of \ac{RFML}, the domain is characterized by the \ac{RF} hardware and channel environments (i.e. \ac{IQ} imbalance, non-linear distortion, \ac{SNR}, multi-path effects), and the task is the application being addressed, including the range of possible outputs (i.e. $n$-class \ac{AMC}, \ac{SEI}, \ac{SNR} estimation).

Recent work presented the \ac{RF}-specific \ac{TL} taxonomy shown in Fig. \ref{fig:taxonomy} \cite{wong2022}, adapted from the general \ac{TL} taxonomy of \cite{pan2010} and the \ac{NLP}-specific taxonomy of \cite{ruder2019}. 
Per this taxonomy, \ac{RF} \ac{TL} is categorized by training data availability and whether or not the source and target tasks differ:
\begin{easylist}[itemize]
@ Domain adaptation is the setting in which source and target tasks are the same, but the source and target domains differ, and can be further categorized as
    @@ Environment adaptation, where the channel environment is changing, but the transmitter/receiver pair(s) are constant, 
    @@ Platform adaptation, where the transmitter/receiver hardware is changing, but the channel environment is consistant, and
    @@ Environment platform co-adaptation, where changes in both the channel environment and transmitter/receiver hardware must be overcome.
@ Multi-task learning is the setting in which different source and target tasks are learned simultaneously.
@ Sequential learning is the setting in which a source task is learned first, and the target task, different from the source task, is learned during a second training phase.
\end{easylist}

Typically, the same training techniques are used to perform both domain adaptation and sequential learning, most commonly head re-training and model fine-tuning, which are the focus of in this work.
Existing works have successfully utilized such techniques to overcome changes in channel environment \cite{chen2019, pati2020} and wireless protocol \cite{kuzdeba2021, robinson2021}, to transfer from synthetic data to captured data \cite{oshea2018, dorner2018, zheng2020, cyborg}, and to add or remove output classes \cite{peng2020}, for a variety of \ac{RFML} use-cases.
Meanwhile, multi-task learning approaches tend to utilize more than one loss term during a single training phase, and has been more commonly used in the context of \ac{ML}-enabled wireless communications systems that use expert-defined features rather than raw \ac{RF} data as input.
However, multi-task learning techniques have been used to facilitate end-to-end communications systems \cite{ye2020}, as well as to improve the explainability and accuracy of \ac{RFML} models \cite{clark2019, wong2021}.
A systematic examination and evaluation of multi-task learning performance is left for future work.

Outside of observing the inability of pre-trained \ac{RFML} models to generalize to new domains/tasks \cite{hauser2018, clark2021, merchant2019}, little-to-no work has examined what characteristics within \ac{RF} data facilitate or restrict transfer \cite{wong2022}.
Without such knowledge, \ac{TL} algorithms for \ac{RFML} are generally restricted to those borrowed from other modalities, such as \ac{CV} and \ac{NLP}.
While correlations can be drawn between the vision or language spaces and the \ac{RF} space, these parallels do not always align, and therefore algorithms designed for \ac{CV} and \ac{NLP} may not always be appropriate for use in \ac{RFML}.
For example, while \ac{CV} algorithm performance is not significantly impacted by a change in the camera(s) used to collect data, so long as the image resolution remains consistent \cite{liu2020}, work in \cite{hauser2018} showed that a change in transmitter/receiver pairs negatively impacted performance by as much as 7\%, despite the collection parameters and even the brand/models of transmitters/receivers remaining consistent.
Therefore, \textit{platform adaptation} techniques that transfer knowledge gleaned from one hardware platform (or set of platforms) to a second hardware platform (or set of platforms) are a necessity in \ac{RFML}, but not in \ac{CV}.

%% file: tables/_signals.tex
\begin{table}[]
\begin{adjustwidth}{-\extralength}{0cm}
\centering
\caption{Signal types included in this work and generation parameters.}
\label{tab:signals}
\begin{tabular}[t]{@{}ll@{}}
\toprule
\multicolumn{1}{c}{\begin{tabular}[c]{@{}c@{}}Modulation\\ Name\end{tabular}} &  \multicolumn{1}{c}{\begin{tabular}[c]{@{}c@{}}Parameter\\ Space\end{tabular}} \\ \midrule 
BPSK & \begin{tabular}[c]{@{}l@{}}Symbol Order \{2\}\\ RRC Pulse Shape\\ Excess Bandwidth \{0.35, 0.5\}\\ Symbol Overlap $\in$ {[}3, 5{]}\end{tabular} \\ \\
QPSK & \begin{tabular}[c]{@{}l@{}}Symbol Order \{4\}\\ RRC Pulse Shape\\ Excess Bandwidth \{0.35, 0.5\}\\ Symbol Overlap $\in$ {[}3, 5{]}\end{tabular} \\ \\
PSK8  & \begin{tabular}[c]{@{}l@{}}Symbol Order \{8\}\\ RRC Pulse Shape\\ Excess Bandwidth \{0.35, 0.5\}\\ Symbol Overlap $\in$ {[}3, 5{]}\end{tabular} \\ \\
PSK16 & \begin{tabular}[c]{@{}l@{}}Symbol Order \{16\}\\ RRC Pulse Shape\\ Excess Bandwidth \{0.35, 0.5\}\\ Symbol Overlap $\in$ {[}3, 5{]}\end{tabular} \\ \\
OQPSK & \begin{tabular}[c]{@{}l@{}}Symbol Order \{4\}\\ RRC Pulse Shape\\ Excess Bandwidth \{0.35, 0.5\}\\ Symbol Overlap $\in$ {[}3, 5{]}\end{tabular} \\ \\
QAM16 & \begin{tabular}[c]{@{}l@{}}Symbol Order \{16\}\\ RRC Pulse Shape\\ Excess Bandwidth \{0.35, 0.5\}\\ Symbol Overlap $\in$ {[}3, 5{]}\end{tabular} \\ \\
QAM32 & \begin{tabular}[c]{@{}l@{}}Symbol Order \{32\}\\ RRC Pulse Shape\\ Excess Bandwidth \{0.35, 0.5\}\\ Symbol Overlap $\in$ {[}3, 5{]}\end{tabular} \\ \\
QAM64 & \begin{tabular}[c]{@{}l@{}}Symbol Order \{64\}\\ RRC Pulse Shape\\ Excess Bandwidth \{0.35, 0.5\}\\ Symbol Overlap $\in$ {[}3, 5{]}\end{tabular} \\ \\
APSK16 & \begin{tabular}[c]{@{}l@{}}Symbol Order \{16\}\\ RRC Pulse Shape\\ Excess Bandwidth \{0.35, 0.5\}\\ Symbol Overlap $\in$ {[}3, 5{]}\end{tabular} \\ \\ \bottomrule
\end{tabular}
\hspace{1cm}
\begin{tabular}[t]{@{}ll@{}}
\toprule
\multicolumn{1}{c}{\begin{tabular}[c]{@{}c@{}}Modulation\\ Name\end{tabular}} & \multicolumn{1}{c}{\begin{tabular}[c]{@{}c@{}}Parameter\\ Space\end{tabular}} \\ \midrule
APSK32 & \begin{tabular}[c]{@{}l@{}}Symbol Order \{32\}\\ RRC Pulse Shape\\ Excess Bandwidth \{0.35, 0.5\}\\ Symbol Overlap $\in$ {[}3, 5{]}\end{tabular} \\ \\
FSK5k & \begin{tabular}[c]{@{}l@{}}Carrier Spacing \{5kHz\}\\ Rect Phase Shape\\ Symbol Overlap \{1\}\end{tabular} \\ \\
FSK75k & \begin{tabular}[c]{@{}l@{}}Carrier Spacing \{75kHz\}\\ Rect Phase Shape\\ Symbol Overlap \{1\}\end{tabular} \\ \\
GFSK5k & \begin{tabular}[c]{@{}l@{}}Carrier Spacing \{5kHz\}\\ Gaussian Phase Shape\\ Symbol Overlap \{2, 3, 4\}\\ Beta $\in$ {[}0.3, 0.5{]}\end{tabular} \\ \\
GFSK75k & \begin{tabular}[c]{@{}l@{}}Carrier Spacing \{75kHz\}\\ Gaussian Phase Shape\\ Symbol Overlap \{2, 3, 4 \}\\ Beta $\in$ {[}0.3, 0.5{]}\end{tabular} \\ \\
MSK & \begin{tabular}[c]{@{}l@{}}Carrier Spacing \{2.5kHz\}\\ Rect Phase Shape\\ Symbol Overlap \{1\}\end{tabular} \\ \\
GMSK & \begin{tabular}[c]{@{}l@{}}Carrier Spacing \{2.5kHz\}\\ Gaussian Phase Shape\\ Symbol Overlap \{2, 3, 4\}\\ Beta $\in$ {[}0.3, 0.5{]}\end{tabular} \\ \\
FM-NB & Modulation Index $\in$ {[}0.05, 0.4{]} \\ \\
FM-WB & Modulation Index $\in$ {[}0.825, 1.88{]} \\ \\
AM-DSB & Modulation Index $\in$ {[}0.5, 0.9{]} \\ \\
AM-DSBSC & Modulation Index $\in$ {[}0.5, 0.9{]} \\ \\
AM-LSB & Modulation Index $\in$ {[}0.5, 0.9{]} \\ \\
AM-USB &  Modulation Index $\in$ {[}0.5, 0.9{]} \\ \\
AWGN &  \\ \bottomrule
\end{tabular}
\end{adjustwidth}
\end{table}

%% file: text/methodology.tex
\begin{figure}[t]
    \begin{adjustwidth}{-\extralength}{0cm}
    \centering
    \begin{subfigure}[b]{.9\textwidth}
      \includegraphics[width=14cm]{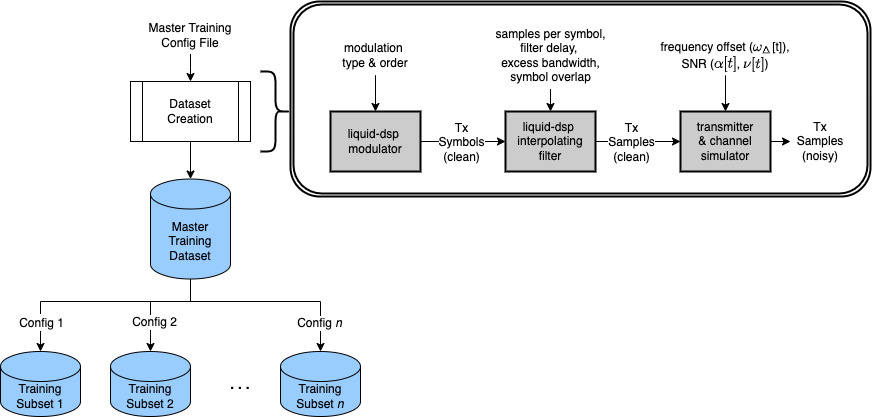}
      \caption{\centering{The training dataset generation process, repeated to create the validation and test datasets.}}
      \vspace{0.5cm}
      \label{fig:sd_data}
    \end{subfigure}
    \begin{subfigure}[b]{.9\textwidth}
      \includegraphics[width=14cm]{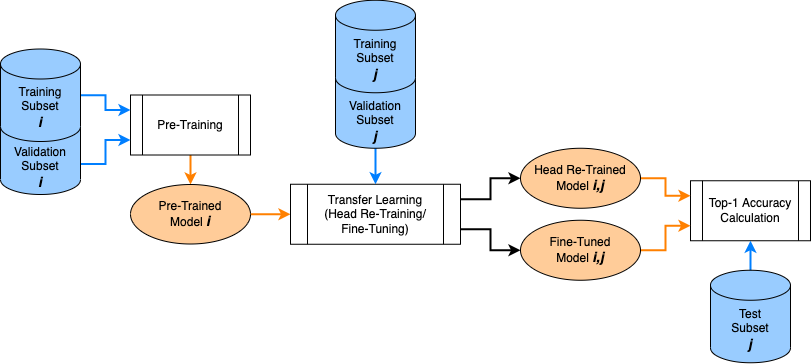}
      \caption{\centering{The process for model pre-training and \ac{TL}.}}
      \label{fig:sd_training}
    \end{subfigure}
    \caption{A system overview of the (a) dataset creation and (b) model pre-training, \ac{TL}, and model evaluation processes used in this work.}
    \label{fig:system_diagram}
    \end{adjustwidth}
\end{figure}

This section presents the experimental setup used in this work, shown in Fig. \ref{fig:system_diagram}, which includes the data and dataset creation process and the model architecture, training, and evaluation, each described in detail in the following subsections. 

\subsection{Dataset Creation}\label{sec:data}
\input{text/data}

\subsection{Model Architecture and Training}\label{sec:training}
\input{text/model}


%% file: text/data.tex
This work used a custom synthetic dataset generation tool based off the open-source signal processing library \textit{liquid-dsp} \cite{liquid}, which allowed for full control over the chosen parameters-of-interest, \ac{SNR}, \ac{FO}, and modulation type, and ensured accurate labelling of the training, validation, and test data.
The dataset creation process, shown in Fig. \ref{fig:sd_data}, begins with the construction of a large ``master" dataset containing all modulation schemes and combinations of \ac{SNR} and \ac{FO} needed for the experiments performed (Section \ref{sec:master}).
Then, for each experiment performed herein, subsets of the data were selected from the master dataset using configuration files containing the desired metadata parameters (Sections \ref{sec:snr} - \ref{sec:modschemes}).
The master dataset is publicly available on IEEE DataPort \cite{wong_dataset_2022}.

\subsubsection{Simulation Environment}\label{sec:sim_env}
All data used in this work was generated using the same noise generation, signal parameters, and signal types as in \cite{clark2019}.
More specifically, in this work, the signal space has been restricted to the 23 signal types shown in Table \ref{tab:signals}, observed at complex baseband in the form of discrete time-series signals, s[t], where
\begin{equation}
s[t] = \alpha_\Delta[t] \cdot \alpha [t] e^{(j\omega[t]+j\theta[t])} \cdot e^{(j\omega_{\Delta}[t]+j\theta_{\Delta}[t])} + \nu[t] \hspace{0.25cm}
\end{equation}
$\alpha[t]$, $\omega[t]$, and $\theta[t]$ are the magnitude, frequency, and phase of the signal at time $t$, and $\nu[t]$ is the additive interference from the channel.
Any values subscripted with a $\Delta$ represent imperfections/offsets caused by the transmitter/receiver and/or synchronization. 
Without loss of generality, all offsets caused by hardware imperfections or lack of synchronization have been consolidated onto the transmitter during simulation.

Signals are initially synthesized in an \ac{AWGN} channel environment with unit channel gain, no phase offset, and frequency offset held constant for each observation.
Like in \cite{clark2019}, \ac{SNR} is defined as
\begin{equation}
\text{SNR} = 10 \log_{10} \bigg( \frac{\sum_{t=0}^{N-1} \lvert s[t] - \nu[t] \rvert ^2}{\sum_{t=0}^{N-1} \lvert \nu[t] \rvert ^2} \bigg)
\end{equation}
where $N$ is the length of the capture measured in samples. 
This definition of \ac{SNR} is based on an oracle-style knowledge of the generated signals, where the symbol energy (Es) has been calibrated relative to its instantaneous noise floor (N0), with the sampling bandwidth being marginally higher than the actual signal bandwidth.
It should be noted that \ac{RFML} approaches generally ingest more than one symbol at a time increasing the effective \ac{SNR}.
Therefore, feature estimation and/or classification is supported at lower \acp{SNR}.

In this work, we assume a blind receiver.
Therefore, no synchronization or demodulation takes place.
As a result, we are not limiting our conclusions by any specific filtering approaches, bandwidths, or other baseband processing.
We do inherently assume all signals are sampled at a sufficiently high rate to meet Nyquist’s sampling theorem.
That is, the \ac{AWGN} captures have a Nyquist rate of 1, and all other captures have a Nyquist rate of either 0.5 or 0.33 (twice or three times the Nyquist bandwidth). 
However, the \ac{AMC} and \ac{TL} approaches used herein do not rely on this critical sampling assumption, as there is no attempt to reconstruct the original signal.

\subsubsection{The Master Dataset}\label{sec:master}
The systematic evaluation of \ac{TL} performance as a function of \ac{SNR}, \ac{FO}, and modulation type conducted in this work is possible through the construction of data-subsets with carefully selected metadata parameters from the larger master dataset.
The constructed master dataset contains 600000 examples of each the signal types given in Table \ref{tab:signals}, for a total of 13.8 million examples.
For each example, the \ac{SNR} is selected uniformly at random between [-10dB, 20dB], the \ac{FO} is selected uniformly at random between [-10\%, 10\%] of the sample rate, and all further signal generation parameters relevant for the signal type, including symbol order, carrier spacing, modulation index, and filtering parameters (excess bandwidth, symbol overlap/filter delay, and/or beta), are selected uniformly at random from the ranges specified in Table \ref{tab:signals}. 
Each example and the associated metadata is saved in SigMF format \cite{hilburn2018}.

\begin{figure}[t]
    \begin{adjustwidth}{-\extralength}{0cm}
    \centering
    \begin{subfigure}[b]{.7\textwidth}
      \centering
      \includegraphics[width=9cm]{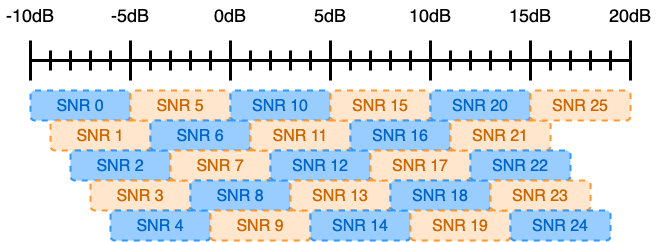}
      \vspace{0.5cm}
      \caption{\centering{Sweep over SNR.}}
      \label{fig:snr_sweep}
      \vspace{0.25cm}
    \end{subfigure}
    \begin{subfigure}[b]{.5\textwidth}
      \centering
      \includegraphics[width=6cm]{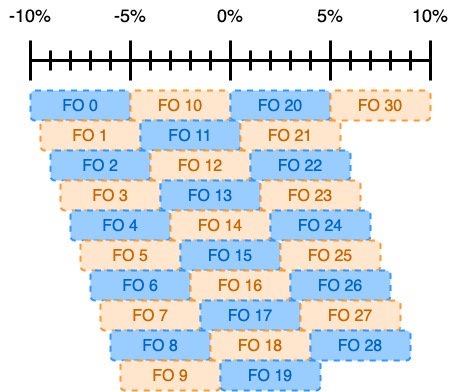}
      \caption{\centering{Sweep over FO.}}
      \label{fig:fo_sweep}
      \vspace{0.25cm}
    \end{subfigure}
    \begin{subfigure}[b]{.9\textwidth}
      \centering
      \includegraphics[width=10cm]{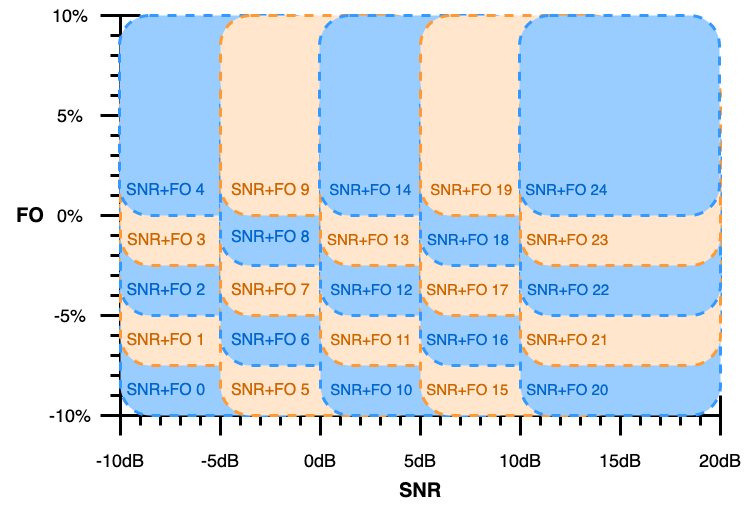}
      \caption{\centering{Sweep over SNR and FO.}}
      \label{fig:snr_fo_sweep}
    \end{subfigure}
    \caption{The parameter-of-interest range for each domain adaptation data subset.}
    \label{fig:data_sweep}
    \end{adjustwidth}
\end{figure}

\subsubsection{The Sweep over \ac{SNR}}\label{sec:snr}
To analyze the impact of \ac{SNR} alone on \ac{TL} performance, 26 source data-subsets were constructed from the larger master dataset using configuration files, as shown in Fig. \ref{fig:snr_sweep}. 
Each data-subset contains examples with \acp{SNR} selected uniformly at random from a 5dB range sweeping from -10dB to 20dB in 1dB steps (i.e. [-10dB, -5dB], [-9dB, -4dB], ..., [15dB, 20dB]), and for each data-subset in this \ac{SNR} sweep, \ac{FO} was selected uniformly at random between [-5\%, 5\%] of sample rate.
This \ac{SNR} sweep yielded 26 pre-trained source models, each of which was transferred to the remaining 25 target data-subsets (as shown in Fig. \ref{fig:sd_training}), yielding 650 models transferred using head re-training and 650 models transferred using fine-tuning.
Additionally, 26 baseline models were trained, as described further in Section \ref{sec:training}.

\subsubsection{The Sweep over \ac{FO}}\label{sec:fo}
To analyze the impact of \ac{FO} alone on \ac{TL} performance, 31 source data-subsets were constructed from the larger master dataset (as shown in Fig. \ref{fig:fo_sweep}) containing examples with \acp{FO} selected uniformly at random from a 5\% range sweeping from -10\% of sample rate to 10\% of sample rate in 0.5\% steps (i.e. [-10\%, -5\%], [-9.5\%, -4.5\%], ..., [5\%, 10\%]).
For each data-subset in this \ac{FO} sweep, \ac{SNR} was selected uniformly at random between [0dB, 20dB].
This \ac{FO} sweep yielded 31 pre-trained source models, each of which was transferred to the remaining 30 target data-subsets (as shown in Fig. \ref{fig:sd_training}) yielding 930 models transferred using head re-training, and 930 models transferred using fine-tuning.
Additionally, 31 baseline models were trained, as described further in Section \ref{sec:training}.

\subsubsection{The Sweep over both \ac{SNR} \& \ac{FO}}\label{sec:snr_fo}
To analyze the impact of both \ac{SNR} and \ac{FO} on \ac{TL} performance, 25 source data-subsets were constructed from the larger master dataset (as shown in Fig. \ref{fig:snr_fo_sweep}) containing examples with \acp{SNR} selected uniformly at random from a 10dB range sweeping from -10dB to 20dB in 5dB steps (i.e. [-10dB, 0dB], [-5dB, 5dB], ..., [10dB, 20dB]) and with \acp{FO} selected uniformly at random from a 10\% range sweeping from -10\% of sample rate to 10\% of sample rate in 2.5\% steps (i.e. [-10\%, 0\%], [-7.5\%, 2.5\%], ..., [0\%, 10\%]).
This \ac{SNR} and \ac{FO} sweep yielded 25 pre-trained source models, each of which was transferred to the remaining 24 target data-subsets (as shown in Fig. \ref{fig:sd_training}) yielding 600 models transferred using head re-training, and 600 models transferred using fine-tuning.
Additionally, 25 baseline models were trained, as described further in Section \ref{sec:training}.

\subsubsection{Modulation Scheme Experiments}\label{sec:modschemes}
To analyze the impact of modulation type on \ac{TL} performance, two groups of data-subsets were constructed.
The first set of data-subsets aims to investigate \ac{TL} performance across broad categories of modulation types, namely linear, frequency-shifted, and analog modulation schemes, as well as datasets containing combinations of modulation types.
More specifically, 5 source data-subsets were constructed from the larger master dataset containing the following modulation schemes:
\begin{easylist}[itemize]
@ All modulations
@ Small Subset -- BPSK, QPSK, OQPSK, QAM16, QAM64, APSK16, FSK 5k, MSK, FM-NB, DSB, USB, AWGN
@ Linear modulations -- BPSK, QPSK, PSK8, PSK16, OQPSK, QAM16, QAM32, QAM64, APSK16, APSK32, AWGN
@ Frequency-shifted modulations -- FSK 5k, FSK 75k, GFSK 5k, GFSK 75k, MSK, GMSK, AWGN
@ Analog modulations -- FM-NB, FM-WB, DSB, DSBSC, LSB, USB, AWGN
\end{easylist}
\noindent For each data-subset in this modulation type experiment, called ``Modulation Experiment 1", \ac{SNR} was selected uniformly at random between [0dB, 20dB] and \ac{FO} was selected uniformly at random between [-5\%, 5\%] of sample rate.
This experiment yielded 5 pre-trained source models, each of which was transferred to the remaining 4 target data-subsets, yielding 20 models transferred using head re-training and 20 models transferred using fine-tuning.
Additionally, 5 baseline models were trained, as described further in Section \ref{sec:training}.

The second set of data-subsets was constructed such that a single modulation type was added/removed from the small/all modulations datasets described above, mimicking a successive model refinement scenario.
More specifically, the 12 source data-subsets were constructed from the larger master dataset containing:
\begin{easylist}[itemize]
@ small -- BPSK, QPSK, OQPSK, QAM16, 64qam, APSK16, FSK 5k, MSK, FM-NB, DSB, USB, AWGN
@ subset1 -- small + PSK8
@ subset2 -- subset1 + PSK16
@ subset3 -- subset2 + QAM32
@ subset4 -- subset3 + APSK32
@ subset5 -- subset4 + FSK 75k
@ subset6 -- subset5 + GFSK 5k
@ subset7 -- subset6 + GFSK 75k
@ subset8 -- subset7 + GMSK
@ subset9 -- subset8 + FM-WB
@ subset10 - subset9 + DSBSC
@ all -- subset10 + LSB
\end{easylist}
Again, \ac{SNR} was selected uniformly at random between [0dB, 20dB] and \ac{FO} was selected uniformly at random between [-5\%, 5\%] of sample rate.
This experiment is called ``Modulation Experiment 2" herein.
This experiment yielded 12 pre-trained source models, each of which was transferred to the remaining 11 target data-subsets, yielding 132 models transferred using head re-training and 132 models transferred using fine-tuning.
Additionally, 12 baseline models were trained, as described further in Section \ref{sec:training}.

%% file: text/model.tex
In this work, we utilize a single architecture trained across pairwise combinations of source/target datasets with varying 
(1) \acp{SNR}, 
(2) \acp{FO}, 
(3) \acp{SNR} and \ac{FO}, or
(4) modulation types
in order to identify the impact of these parameters-of-interest on \ac{TL} performance. 
Given the large number of models trained for this work, training time was a primary concern when selecting the model architecture. 
Therefore, this work uses a simple \ac{CNN} architecture, shown in Table \ref{tab:model}, that is based off of the architectures used in \cite{clark2019} and \cite{wong2021cb}, with a reduction in the input size. 
Although many works including \cite{clark2019} and \cite{wong2021cb} have found success using larger input sequences, works such as \cite{oshea2016} and \cite{west2017} have found 128 input samples to be sufficient.
Recognizing that longer input sequences results in increased computation and training time, in this work, 128 raw \ac{IQ} samples are used as input corresponding to approximately 16-32 symbols depending on the symbol rate of the example.
These samples are fed to the network in a $(1, 2, 128)$ tensor, such that 1 refers to the number of channels, 2 refers to the \ac{IQ} components, and 128 refers to the number of samples.
The network contains two 2D convolutional layers, the first uses 1500 kernels of size $(1, 7)$ and the second uses 260 kernels of size $(2, 7)$.
The second convolutional layer is followed by a flattening layer, a dropout layer using a rate of 0.5, and two linear fully-connected layers containing 65 and $n$ nodes where $n$ is the number of output classes (i.e. modulation schemes) being trained.
Both convolutional layers and the first linear layer use a ReLU activation function, and the final linear layer uses a Softmax activation function.

\input{tables/model}

The model pre-training and \ac{TL} process is shown in Fig. \ref{fig:sd_training}, and represents a standard training pipeline. 
For pre-training, the training dataset contained 5000 examples per class, and the validation dataset contained 500 examples per class.
These dataset sizes are consistent with \cite{clark2019} and adequate to achieve consistent convergence. 
Each model was trained using the Adam optimizer \cite{kingma2014} and Cross Entropy Loss \cite{pytorchCE}, with the PyTorch default hyper-parameters \cite{pytorch} (a learning rate of 0.001, without weight decay), for a total of 100 epochs. 
A checkpoint was saved after the epoch with the lowest validation loss, and was reloaded at the conclusion of the 100 epochs.

This work examines both head re-training and model fine-tuning methods. 
For head re-training and model fine-tuning, the training dataset contained 500 examples per class, and the validation dataset contained 50 examples per class, representing a smaller sample of available target data.
The head re-training and fine-tuning processes both used the Adam optimizer and Cross Entropy Loss as well, with checkpoints saved at the lowest validation loss.
During head re-training, only the final layer of the model was trained, again using the PyTorch default hyper-parameters, while the rest of the model's parameters were frozen.
During fine-tuning, the entire model was trained with a learning rate of 0.0001, an order of magnitude smaller than the PyTorch default of 0.001.
Finally, all baseline models were trained using the same training process as the pre-trained models, but with 500 training examples per class and 50 validation examples per class, as was used in the \ac{TL} setting.

%% file: tables/model.tex

\begin{table}[]
\centering
\caption{Model architecture with \textit{n} being the number of output classes (modulation types) trained.}
\label{tab:model}
\begin{tabular}{@{}lcc@{}}
\toprule
Layer Type & Num Kernels/Nodes & Kernel Size \\ \midrule
Input & size = (2, 128) &  \\
Conv2d & 1500 & (1, 7) \\
ReLU &  &  \\
Conv2d & 260 & (2, 7) \\
ReLU &  &  \\
Dropout & rate = 0.5 &  \\
Flatten &  &  \\
Linear & 65 &  \\
ReLU &  &  \\
Linear & \textit{n} &  \\ 
Softmax &  &  \\ \midrule
\multicolumn{3}{c}{Trainable Parameters: $7432725 + (66 \cdot n)$} \\ \bottomrule
\end{tabular}
\end{table}

%% file: text/results.tex
The product of the experiments performed herein is 98 data subsets, each with distinct \ac{RF} domains and tasks, corresponding baseline and source models trained from random initialization, and 4664 transfer learned models, half transferred using head re-training and the remaining half transferred using fine-tuning.
Given the careful curation of the signal parameters contained within each data subset, as well as the breadth of signal types and parameters observed, generalized conclusions can be drawn regarding \ac{TL} performance as a function of changes in the propagation environment (\ac{SNR}), transmitter/receiver hardware (\ac{FO}), and \ac{AMC} task.
However, it should be noted that further experiments using captured data are required in order to draw more concrete guidelines for using \ac{RF} \ac{TL} in the field \cite{clark2020}, and is left for future work.
The following subsections present the results obtained from the experiments performed, and discuss insights and practical takeaways that can be gleaned from the results given.

\begin{figure}[t]
    \centering
    \includegraphics[width=0.7\linewidth]{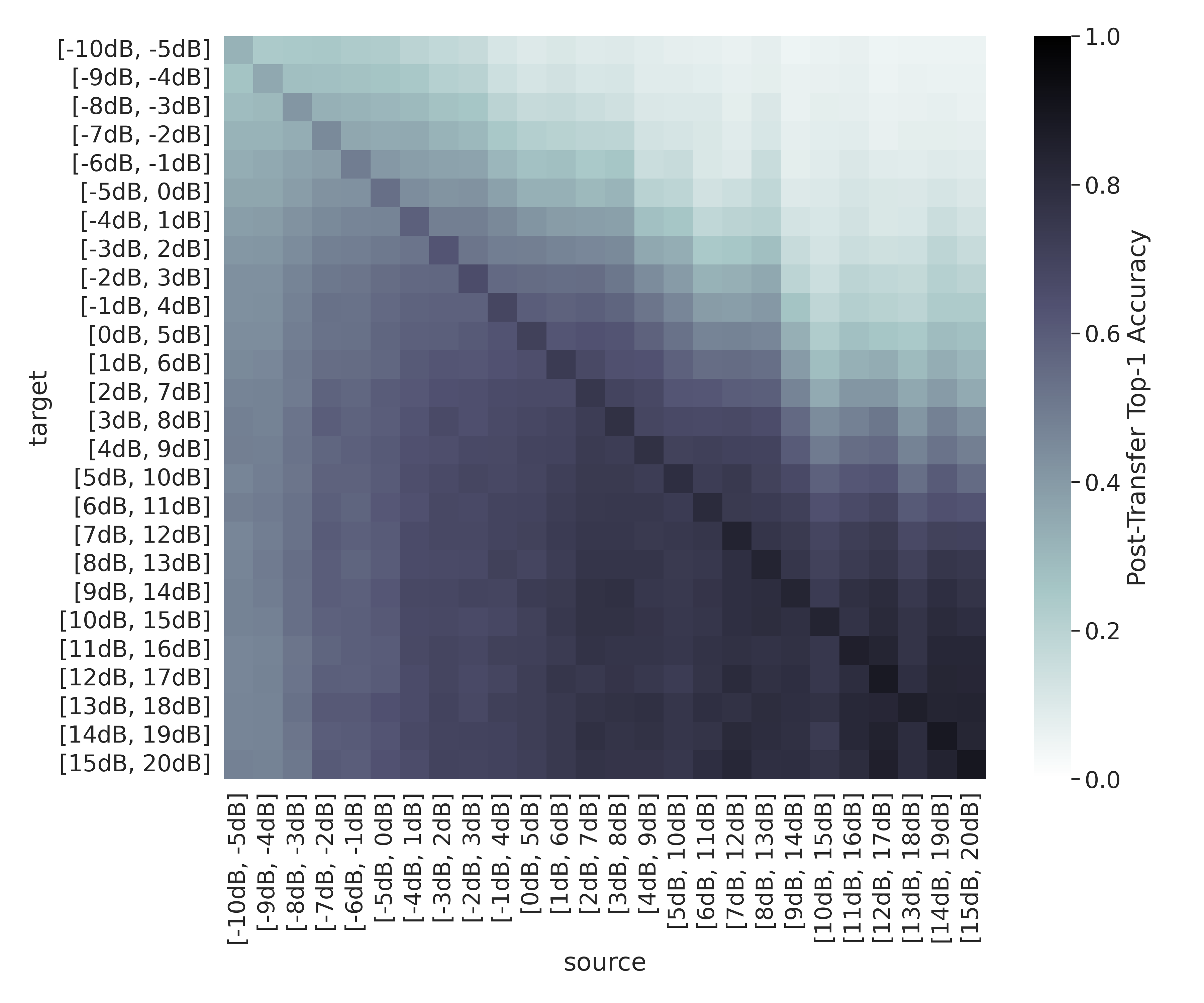}
    \caption{The post-transfer top-1 accuracy for each source/target dataset pair constructed for the sweep over \ac{SNR} using head re-training to perform domain adaptation, shown on a scale of [0.0, 1.0]. When fine-tuning is used to perform domain adaptation, the same trends are apparent.}
    \label{fig:snr_heatmap}
\end{figure}
\unskip

\begin{figure}[h!]
    \centering
    \includegraphics[width=0.7\linewidth]{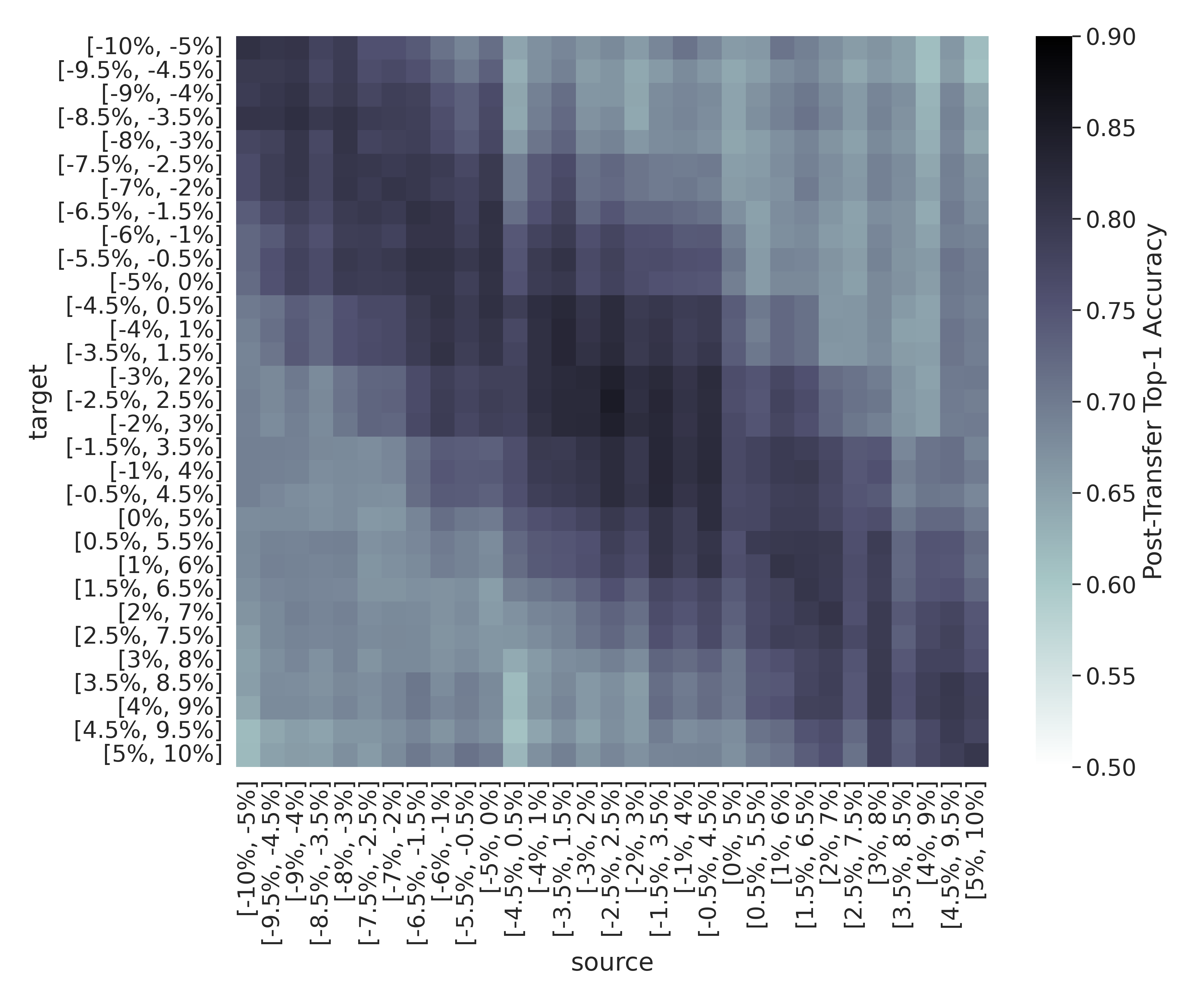}
    \caption{The post-transfer top-1 accuracy for each source/target dataset pair constructed for the sweep over \ac{FO} using head re-training to perform domain adaptation, shown on a scale of [0.5, 0.9]. When fine-tuning is used to perform domain adaptation, the same trends are apparent.}
    \label{fig:fo_heatmap}
\end{figure}
\unskip

\begin{figure}[h]
    \centering
    \includegraphics[width=0.8\linewidth]{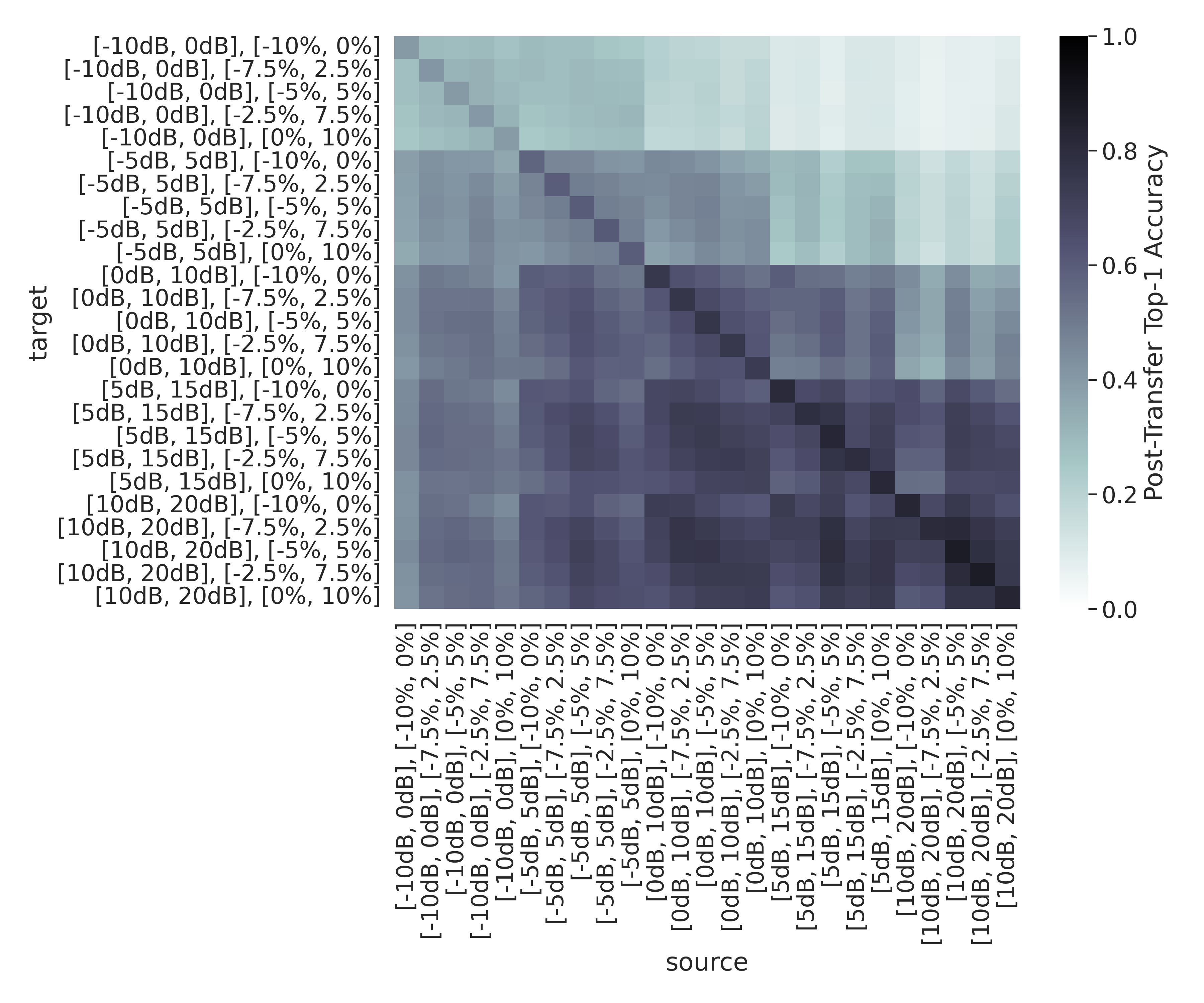}
    \caption{The post-transfer top-1 accuracy for each source/target dataset pair constructed for the sweep over both \ac{SNR} and \ac{FO} using head re-training to perform domain adaptation, shown on a scale of [0.0, 1.0]. When fine-tuning is used to perform domain adaptation, the same trends are apparent.}
    \label{fig:snr_fo_heatmap}
\end{figure}
\unskip

\subsection{When and How is \ac{RF} Domain Adaptation most successful?}
\subsubsection{Impact of Source/Target Domain Similarity and ``Difficulty" on Transfer Performance}
The heatmaps in Figs. \ref{fig:snr_heatmap}-\ref{fig:snr_fo_heatmap} show the post-transfer top-1 accuracy achieved with each of the source/target pairs.
Note that the post-transfer top-1 accuracy results shown in Figs. \ref{fig:snr_heatmap}-\ref{fig:snr_fo_heatmap} are from the models that used head re-training to transfer from the source to target domains/datasets.
However, the accuracy results from the models that used fine-tuning for transfer show the same trends.

Figs. \ref{fig:snr_heatmap}-\ref{fig:snr_fo_heatmap} show that highest post-transfer performance is achieved along the diagonal of the heatmap, where the source and target domains are most similar.
It should be noted that while the notion of domain similarity is ill-defined in general, for the purposes of this work, we are able to say that domains are more similar when the difference between the source and target \ac{SNR} and/or \ac{FO} ranges is smaller, as all other data generation parameters are held constant.
These trends are expected, as models trained on similar domains likely learn similar features, and is consistent with the general theory of \ac{TL} \cite{pan2010}, as well as existing works in modalities outside of \ac{RF} \cite{rosenstein2005}.

Figs. \ref{fig:snr_heatmap}-\ref{fig:snr_fo_heatmap} also show that transfer across changes in \ac{FO} is approximately symmetric, while transfer across changes in \ac{SNR} are not.
This behavior is also expected, and can be attributed to changes in the relative ``difficulty" between the source and target domains.
More specifically, changing the source/target \ac{SNR} inherently changes the difficulty of the problem, as performing \ac{AMC} in lower \ac{SNR} channel environments is more challenging than performing \ac{AMC} in high \ac{SNR} channel environments.
Therefore, the source models trained on the lower \ac{SNR} ranges will transfer to the higher \ac{SNR} ranges, though may not perform optimally, while the source models trained on the higher \ac{SNR} ranges will fail to transfer to the lower \ac{SNR} ranges, as shown in Figs. \ref{fig:snr_heatmap} and \ref{fig:snr_fo_heatmap}.
In contrast, changing the source/target \ac{FO} does not make performing \ac{AMC} any more or less difficult, but may require modifications to the learned features to accommodate which can be likened to performing \ac{FO} calibration, as is standard practice in \ac{RF} receiver operations.
Consequently, small changes in \ac{FO}, $\omega_\Delta[t]$, in either the positive and negative direction, are expected to perform similarly.
Figure \ref{fig:fo_heatmap} indeed shows that \ac{TL} performance is approximately symmetric, with best performance closest to the diagonal where the source and target \ac{FO} ranges are most closesly aligned.

Practically, these trends indicate that the effectiveness of \ac{RF} domain adaptation increases as the source and target domains become more and more similar, and, when applicable, \ac{RF} domain adaptation is more often successful when transferring from harder to easier domains when compared to transferring from easier to harder domains.
For example, transferring from [-5dB, 0dB] to [0dB, 5dB] \ac{SNR} is likely more effective than transferring from [5dB, 10dB] to [0dB, 5dB] \ac{SNR} because although the similarity of between datasets in these two transfer scenarios is the same, [-5dB, 0dB] is a more challenging domain than [0dB, 5dB] whereas [5dB, 10dB] is an easier domain than [0dB, 5dB].
However, transferring from a \ac{FO} range of [-9\%, -4\%] of sample rate to [-8\%, -3\%] of sample rate is likely more effective than transferring from a \ac{FO} range of [-10\%, -5\%] of sample rate to [-8\%, -3\%] of sample rate because [-9\%, -4\%] of sample rate and [-8\%, -3\%] of sample rate are more similar than [-10\%, -5\%] of sample rate and [-8\%, -3\%] of sample rate.

\begin{figure}[t]
    \centering
      \centering
      \includegraphics[width=0.7\linewidth]{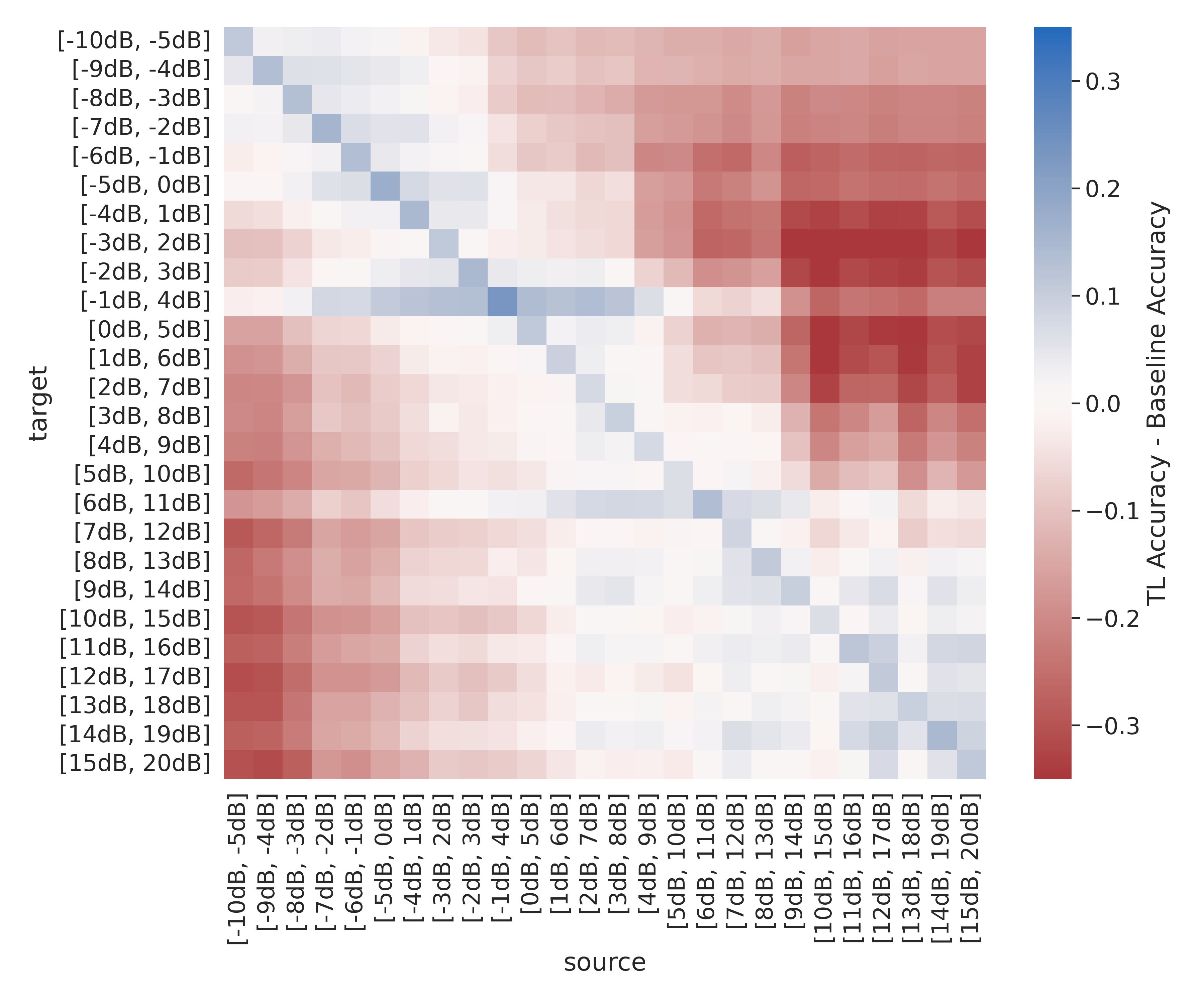}
      \caption{The difference between post-transfer top-1 accuracy and target baseline accuracy for the sweep over SNR using head re-training, shown on a scale of [-0.35, 0.35]. When fine-tuning is used to perform domain adaptation, the same trends are apparent. Note that the increase in performance when the source and target are the same (i.e. along the diagonal) is due to the 10x increase in training data between the baseline and pre-trained models.}
      \label{fig:baseline_diff_snr}
\end{figure}
\unskip

\begin{figure}[h!]
      \centering
      \includegraphics[width=0.7\linewidth]{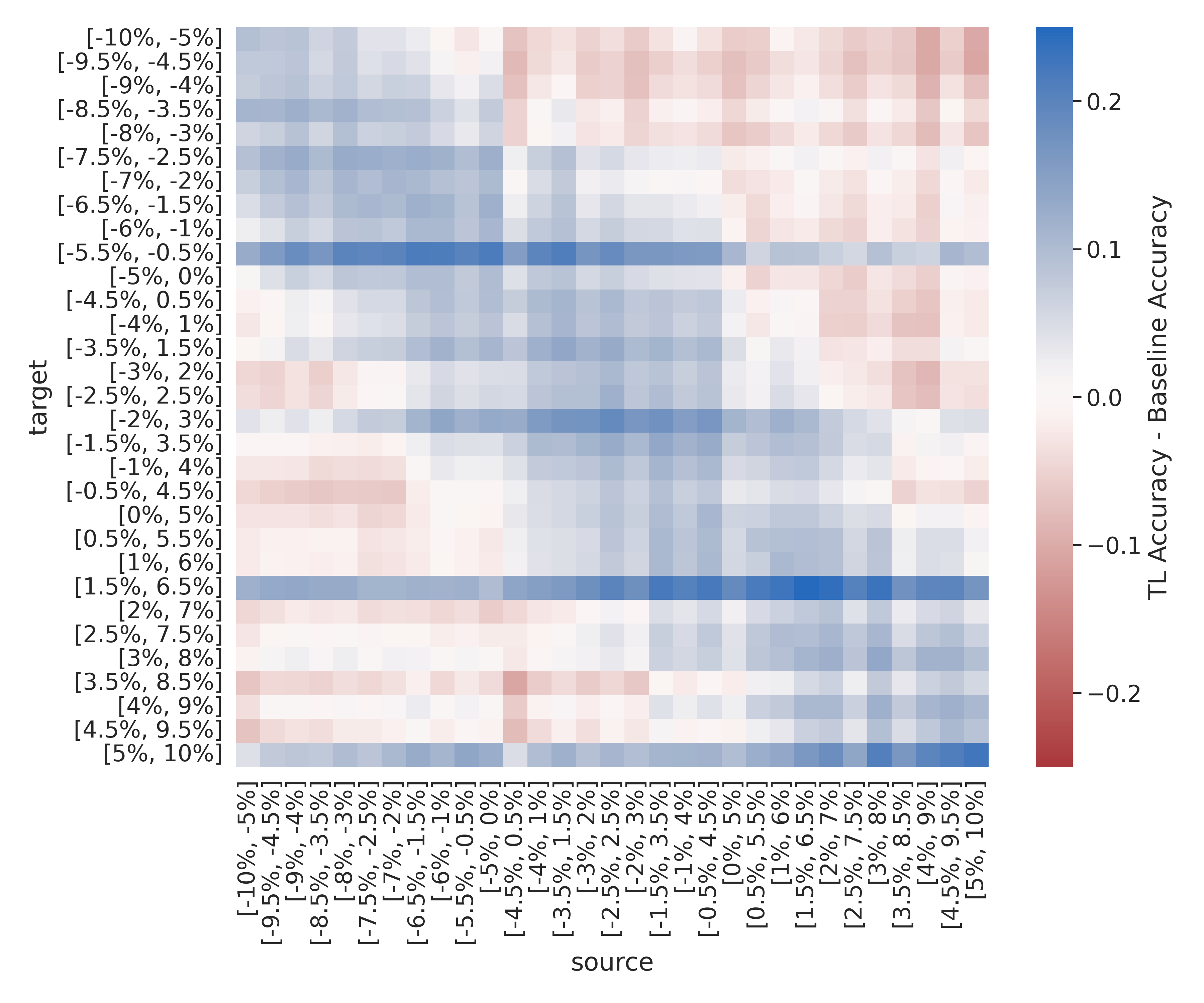}
      \caption{The difference between post-transfer top-1 accuracy and target baseline accuracy for the sweep over FO using head re-training, shown on a scale on [-0.25, 0.25]. When fine-tuning is used to perform domain adaptation, the same trends are apparent. Note that the increase in performance when the source and target are the same (i.e. along the diagonal) is due to the 10x increase in training data between the baseline and pre-trained models.}
    \label{fig:baseline_diff_fo}
\end{figure}
\unskip

\subsubsection{Environment Adaptation vs. Platform Adaptation}
Recalling that the sweep over \ac{SNR} can be regarded as an environment adaptation experiment and the sweep over \ac{FO} can be regarded as a platform adaptation experiment, more general conclusions can be drawn regarding the challenges that environment and platform adaptation present.
From the discussion in the previous subsection regarding the impact that \ac{SNR} and \ac{FO} have on the relative difficulty of the \ac{AMC} task, it follows that changes in \ac{SNR} are likely more challenging to overcome than changes in \ac{FO}.
That is, changes in channel environment are more challenging to overcome using \ac{TL} techniques than changes in transmitter/receiver hardware, such that environment adaptation is more difficult to achieve than platform adaptation
While this trend is indirectly shown through the range of accuracies achieved in Figs. \ref{fig:snr_heatmap}-\ref{fig:snr_fo_heatmap}, which is smaller for the \ac{FO} sweep than the \ac{SNR} sweep and \ac{SNR} $+$ \ac{FO} sweep, and is more directly shown in Figs. \ref{fig:baseline_diff_snr} and \ref{fig:baseline_diff_fo}.

Figs. \ref{fig:baseline_diff_snr} and \ref{fig:baseline_diff_fo} present the difference between post-transfer top-1 accuracy and target baseline accuracy for the \ac{SNR} and \ac{FO} sweeps, such that when the difference value is positive the \ac{TL} model outperforms the baseline model and vice versa.
These results show that for the sweep over \ac{SNR}, the \ac{TL} model only outperforms the baseline near the diagonal where the source and target are very similar.
However, for the sweep over \ac{FO}, the \ac{TL} model outperforms the baseline for a greater number of source/target pairs.
 
From these results, we can conclude that in practice \ac{TL} is more useful for overcoming platform discrepancies than channel discrepancies, unless the channel discrepancy is small.
If the channel discrepancy between source and target is large, one might consider simply training for random initialization on the target data to achieve top performance.
Furthermore, if choosing a source dataset/model for a given target domain/task, one should consider the similarity of of the source/target channel environment before the similarity of the source/target platform, as changes in transmitter/receiver pair are more easily overcome during \ac{TL}.

\begin{figure}[t]
    \centering
    \includegraphics[width=0.7\linewidth]{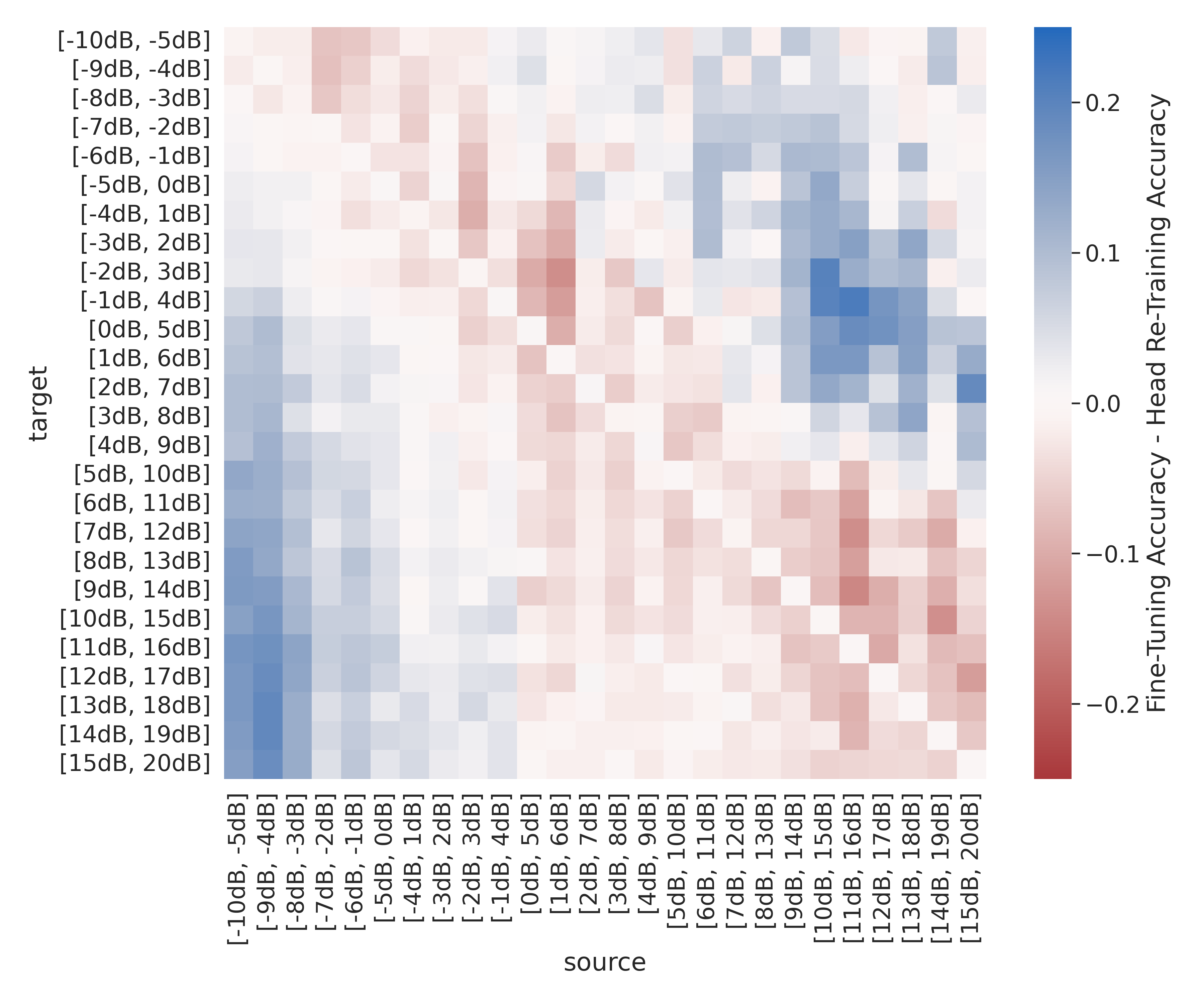}
    \caption{The difference between post-transfer top-1 accuracies achieved using head re-training versus fine-tuning for the sweep over SNR, shown on a scale of [-0.25, 0.25]. When the value is positive, fine-tuning outperforms head re-training. When the value is negative, head re-training outperforms fine-tuning.}
    \label{fig:snr_method_diff}
\end{figure}
\unskip

\begin{figure}[h!]
    \centering
    \includegraphics[width=0.7\linewidth]{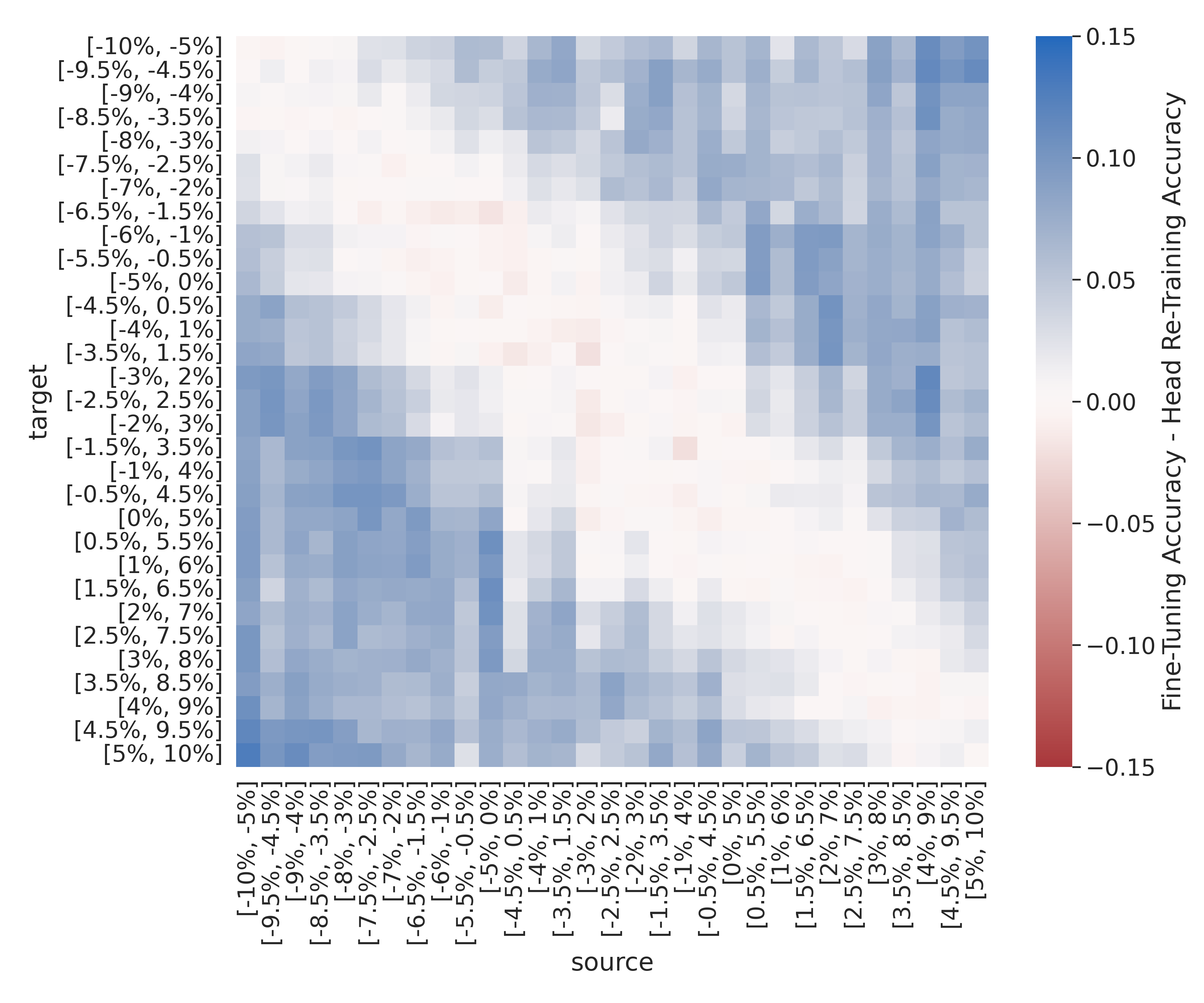}
    \caption{The difference between post-transfer top-1 accuracies achieved using head re-training versus fine-tuning for the sweep over FO, shown on a scale of [-0.15, 0.15]. When the value is positive, fine-tuning outperforms head re-training. When the value is negative, head re-training outperforms fine-tuning.}
    \label{fig:fo_method_diff}
\end{figure}
\unskip

\begin{figure}[h!]
    \centering
    \includegraphics[width=0.8\linewidth]{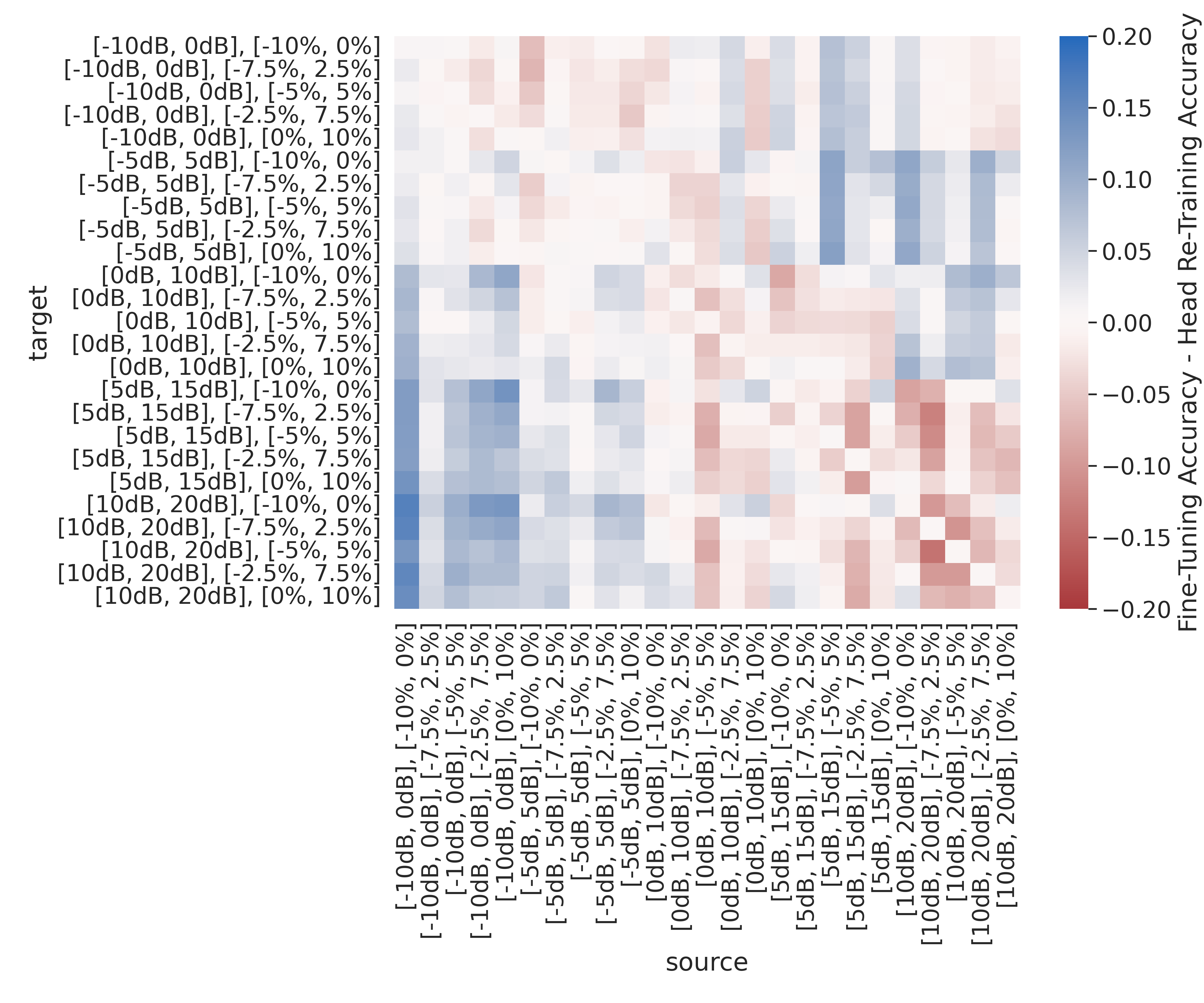}
    \caption{The difference between post-transfer top-1 accuracies achieved using head re-training versus fine-tuning for the sweep over SNR and FO, shown on a scale of [-0.2, 0.2]. When the value is positive, fine-tuning outperforms head re-training. When the value is negative, head re-training outperforms fine-tuning.}
    \label{fig:snr_fo_method_diff}
\end{figure}
\unskip

\subsubsection{Head Re-Training vs. Fine-Tuning}\label{sec:da_method_comparison}
Figs. \ref{fig:snr_method_diff} - \ref{fig:snr_fo_method_diff} plot the difference between post-transfer top-1 accuracies achieved using head re-training versus fine-tuning such that positive values correspond to better fine-tuning performance and negative values correspond to better head re-training performance.
These figures indicate that head re-training is as effective, if not more effective, than fine-tuning when the source and target domains are similar.
Meanwhile, fine-tuning is more effective when the source and target domains are more dissimilar.
Intuitively, this means that when the source/target domains are dissimilar, the features found in the early layers of the source model needed modification to discern between modulation types in the target domain.
However, recalling that Figs. \ref{fig:baseline_diff_snr} and \ref{fig:baseline_diff_fo} showed \ac{TL} only provides benefit when the source and target are somewhat similar, we can conclude that head re-training is as effective, if not more effective, than fine-tuning in the settings where \ac{TL} increases performance over the baseline.
Given that head re-training is more time efficient and less computationally expensive than fine-tuning, there is a strong case for using head re-training over fine-tuning when performing \ac{RF} domain adaptation. 

\begin{figure}[t]
    \centering
    \includegraphics[width=0.7\linewidth]{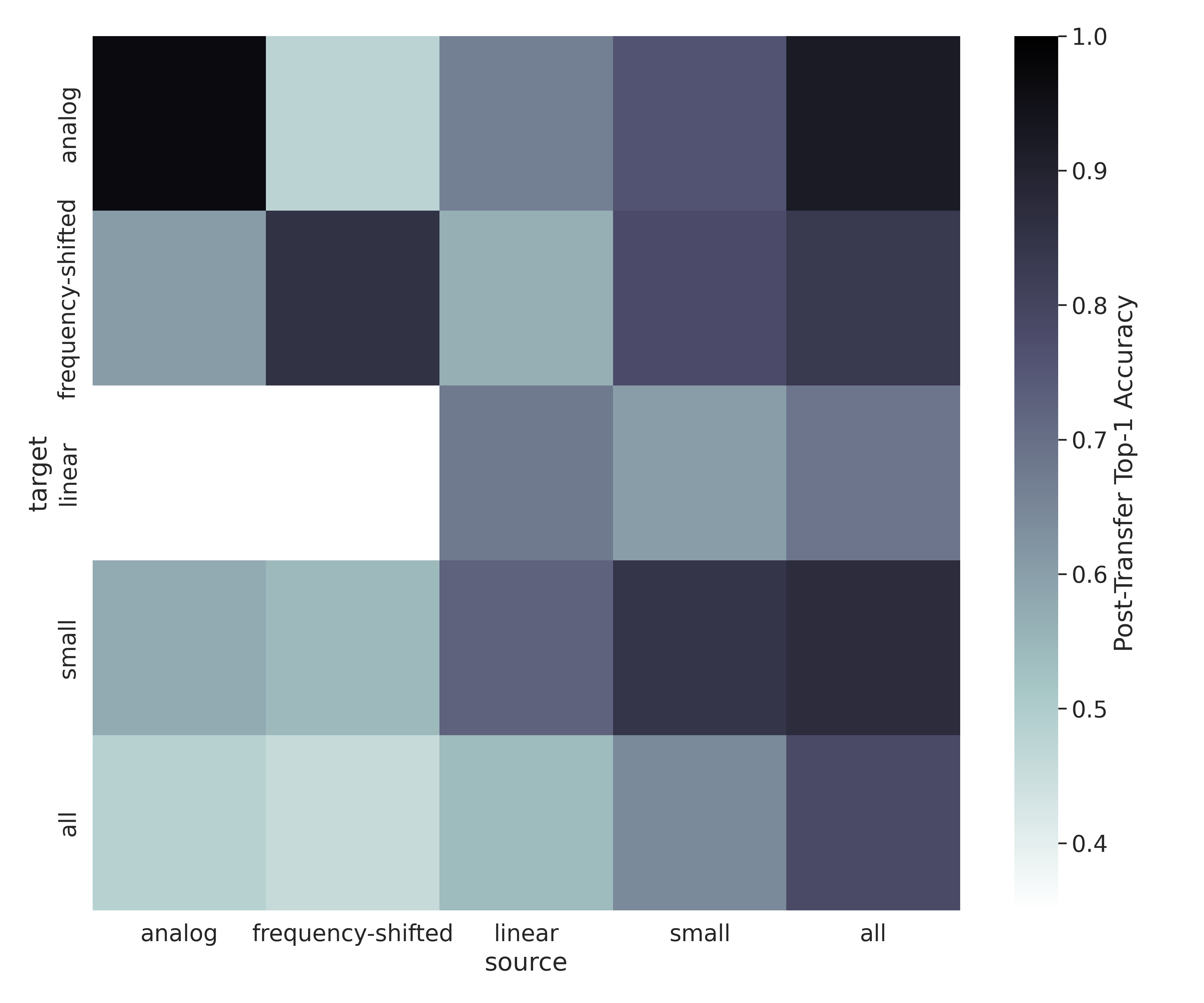}
    \caption{The post-transfer top-1 accuracy for each source/target dataset pair constructed for Modulation Experiment 1 using head re-training, shown on a scale of [0.35, 1.0]. When fine-tuning is used, the same trends are apparent.}
    \label{fig:modgroup1_heatmap}
\end{figure}
\unskip

\begin{figure}[h!]
    \centering
    \includegraphics[width=0.7\linewidth]{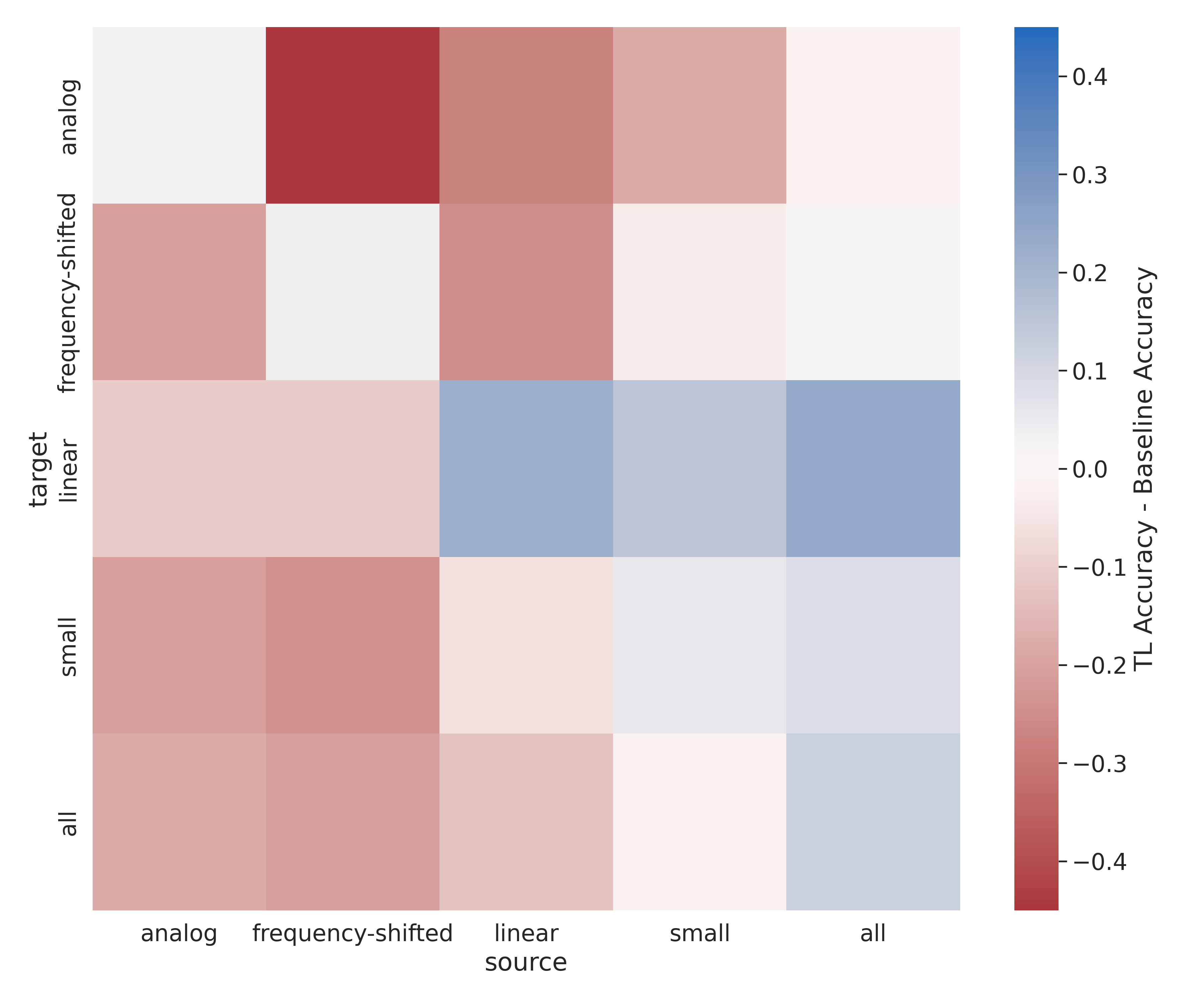}
    \caption{The difference between post-transfer top-1 accuracy and target baseline accuracy for Modulation Experiment 1 using head re-training, shown on a scale of [-0.45, 0.45]. When fine-tuning is used, the same trends are apparent. Note that the increase in performance when the source and target are the same (i.e. along the diagonal) is due to the 10x increase in training data between the baseline and pre-trained models.}
    \label{fig:baseline_diff_modgroup1}
\end{figure}
\unskip

\subsection{When and How is RF Sequential Learning Most Successful?}
\subsubsection{Sequential Learning Across Signal Types: Modulation Experiment 1}
Fig. \ref{fig:modgroup1_heatmap} shows the post-transfer top-1 accuracy for each source/target dataset in Modulation Experiment 1.
As in the domain adaptation experiments discussed previously, the best transfer generally occurs along the diagonal of the heatmap, where the source/target similarity is highest. 
Additionally, the subsets containing only a single type of modulation scheme (i.e. the analog, frequency-shifted, linear subsets) don't transfer well between one another, and also don't transfer well to the subsets which contain multiple types of modulation schemes (i.e. the small and all subsets).
Meanwhile, the small and all subsets transfer fairly well to the analog, frequency-shifted, and linear subsets. 
These results are verified by the results shown in Fig. \ref{fig:baseline_diff_modgroup1} which presents the difference between post-transfer top-1 accuracy and the baseline target models, and shows that \ac{TL} only increases performance over the baseline models when there is significant overlap between the modulation schemes in each subset.

These results are expected when we consider the general setting in which \ac{TL} is beneficial: when the source and target are ``similar".
When no similar signal types between source/target there is little-to-no benefit to using \ac{TL}, such as when attempting transfer between the analog, frequency-shifted, and linear subsets.
However, because the small and all subsets contain at least one modulation scheme from each of the analog, frequency-shifted, and linear subsets, the pre-trained source model has some prior knowledge of each category of modulation schemes from which to build.
Practically, these results indicate that \ac{TL} is only beneficial when similar signal types in the source and target datasets.

\begin{figure}[t]
    \centering
    \includegraphics[width=0.7\linewidth]{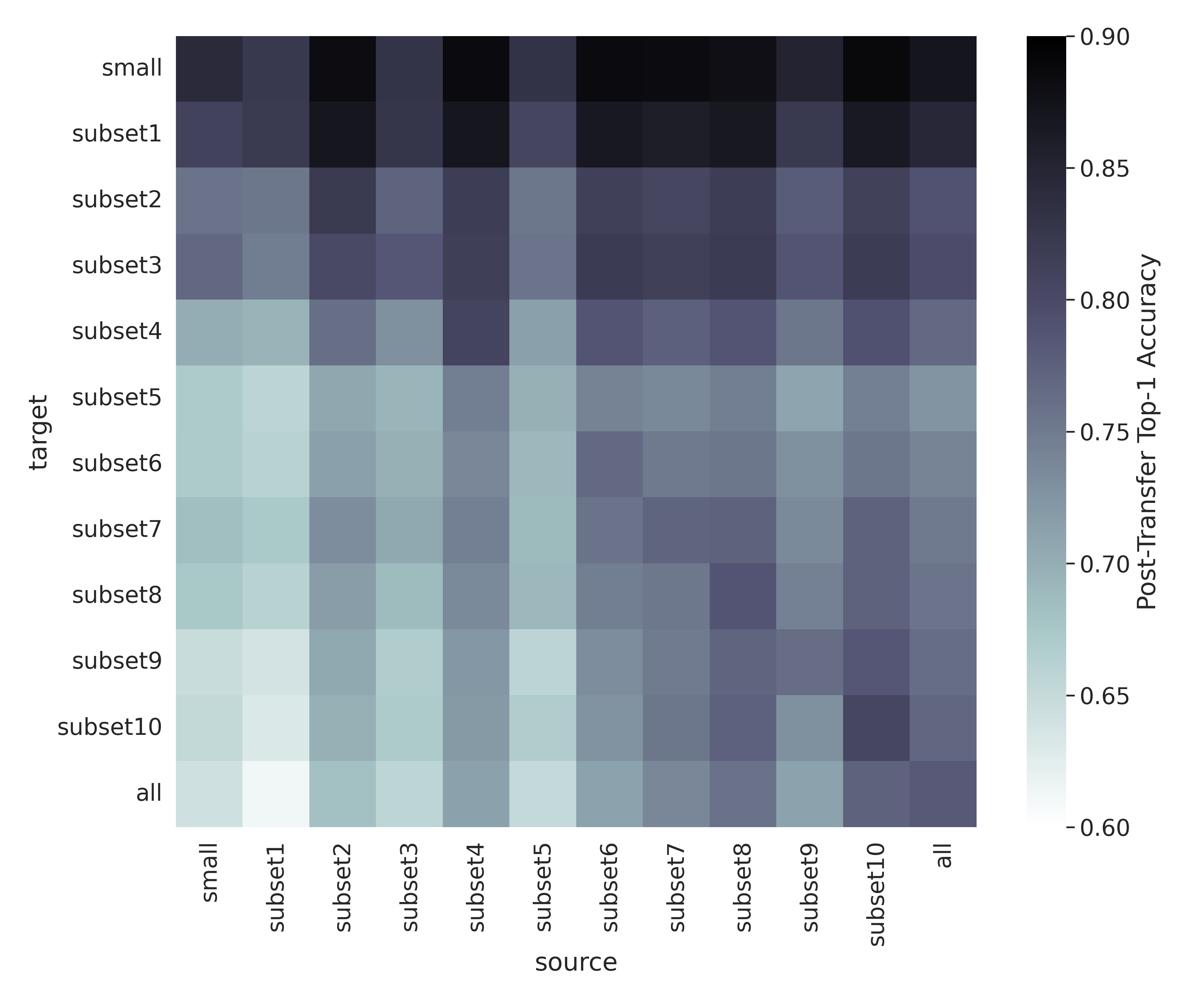}
    \caption{The post-transfer top-1 accuracy for each source/target dataset pair constructed for Modulation Experiment 2 using head re-training, shown on a scale of [0.6, 0.9]. When fine-tuning is used, the same trends are apparent.}
    \label{fig:modgroup2_heatmap}
\end{figure}
\unskip

\begin{figure}[h!]
      \centering
      \includegraphics[width=0.8\linewidth]{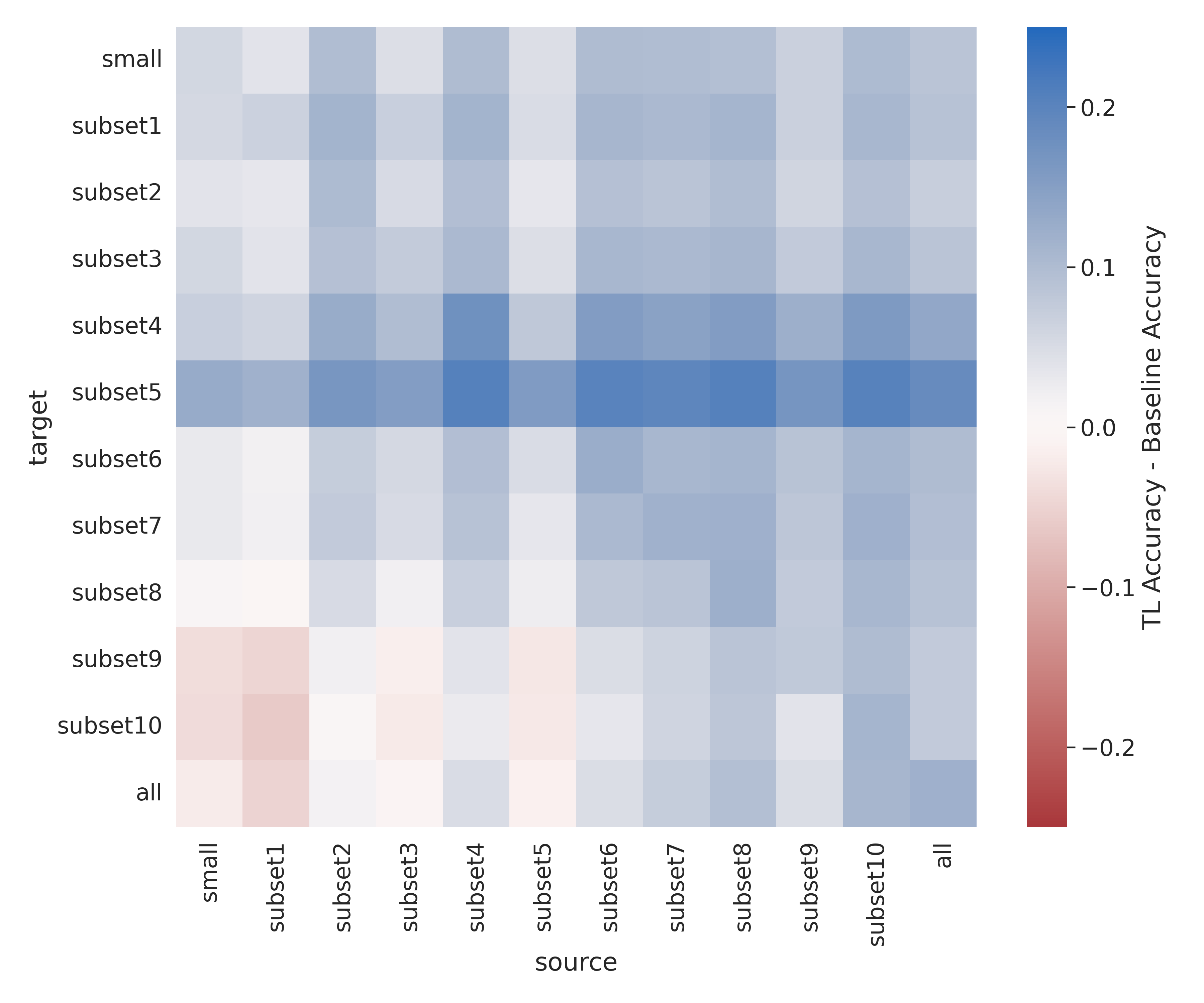}
      \caption{The difference between post-transfer top-1 accuracy and target baseline accuracy for Modulation Experiment 2 using head re-training, shown on a scale of [-0.25, 0.25]. When fine-tuning is used, the same trends are apparent. Note that the increase in performance when the source and target are the same (i.e. along the diagonal) is due to the 10x increase in training data between the baseline and pre-trained models.}
    \label{fig:baseline_diff_modgroup2}
\end{figure}
\unskip

\subsubsection{Sequential Learning For Successive Model Refinement: Modulation Experiment 2}
Figs. \ref{fig:modgroup2_heatmap} and \ref{fig:baseline_diff_modgroup2} present the post-transfer top-1 accuracy and the difference between the post-transfer top-1 accuracy for Modulation Experiment 2 and target baseline accuracy for Modulation Experiment 2 respectively.
These results indicate that it is easier to remove output classes during the \ac{TL} phase than it is to add output classes, as evidenced by higher performance in the upper triangle of the heatmap in Fig. \ref{fig:modgroup2_heatmap}, as well as the significant performance benefits over the target baseline models shown in Fig. \ref{fig:baseline_diff_modgroup2}.
This behavior is expected, as, intuitively, it is easier to forget or disregard prior knowledge than to acquire new knowledge during transfer.
More specifically, by pre-training on a larger subset of signal types (i.e. outputs), the source model has already learned features to identify all of the modulation classes in the target task.
In fact, the source model has likely learned more features than necessary to perform the target task, and could undergo feature pruning in order to reduce computational complexity.
It should also be noted that the task gets easier as output classes are removed, further contributing to the trend.
Practically, these results dictate that one should utilize a source task that encompasses the target task, when possible.

\begin{figure}[h!]
    \centering
    \includegraphics[width=0.7\linewidth]{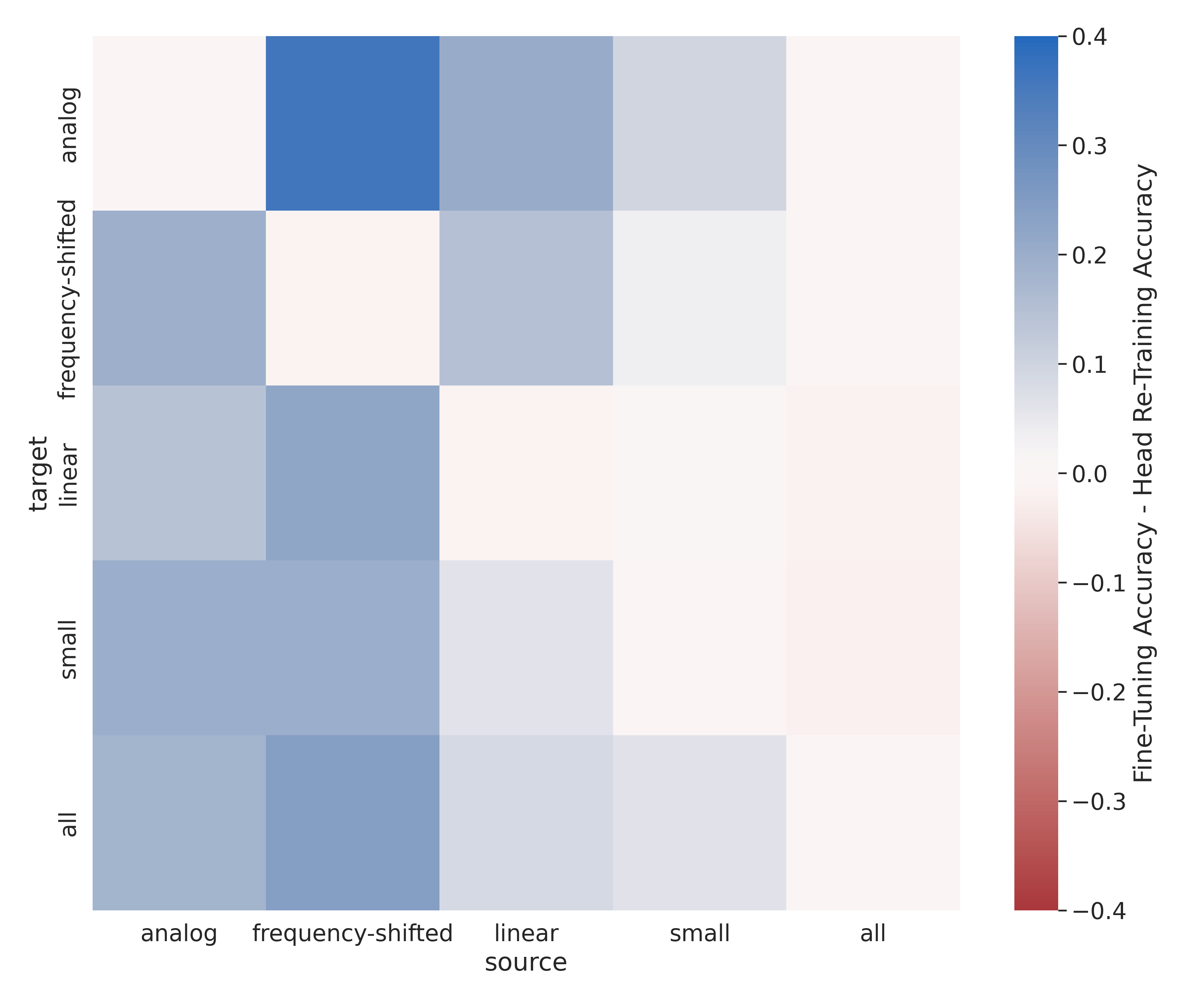}
    \caption{The difference between post-transfer top-1 accuracies achieved using head re-training versus fine-tuning for Modulation Experiment 1, shown on a scale of [-0.4, 0.4]. When the value is positive, fine-tuning outperforms head re-training. When the value is negative, head re-training outperforms fine-tuning.}
    \label{fig:modgroup1_method_diff}
\end{figure}
\unskip

\begin{figure}[h!]
    \centering
    \includegraphics[width=0.8\linewidth]{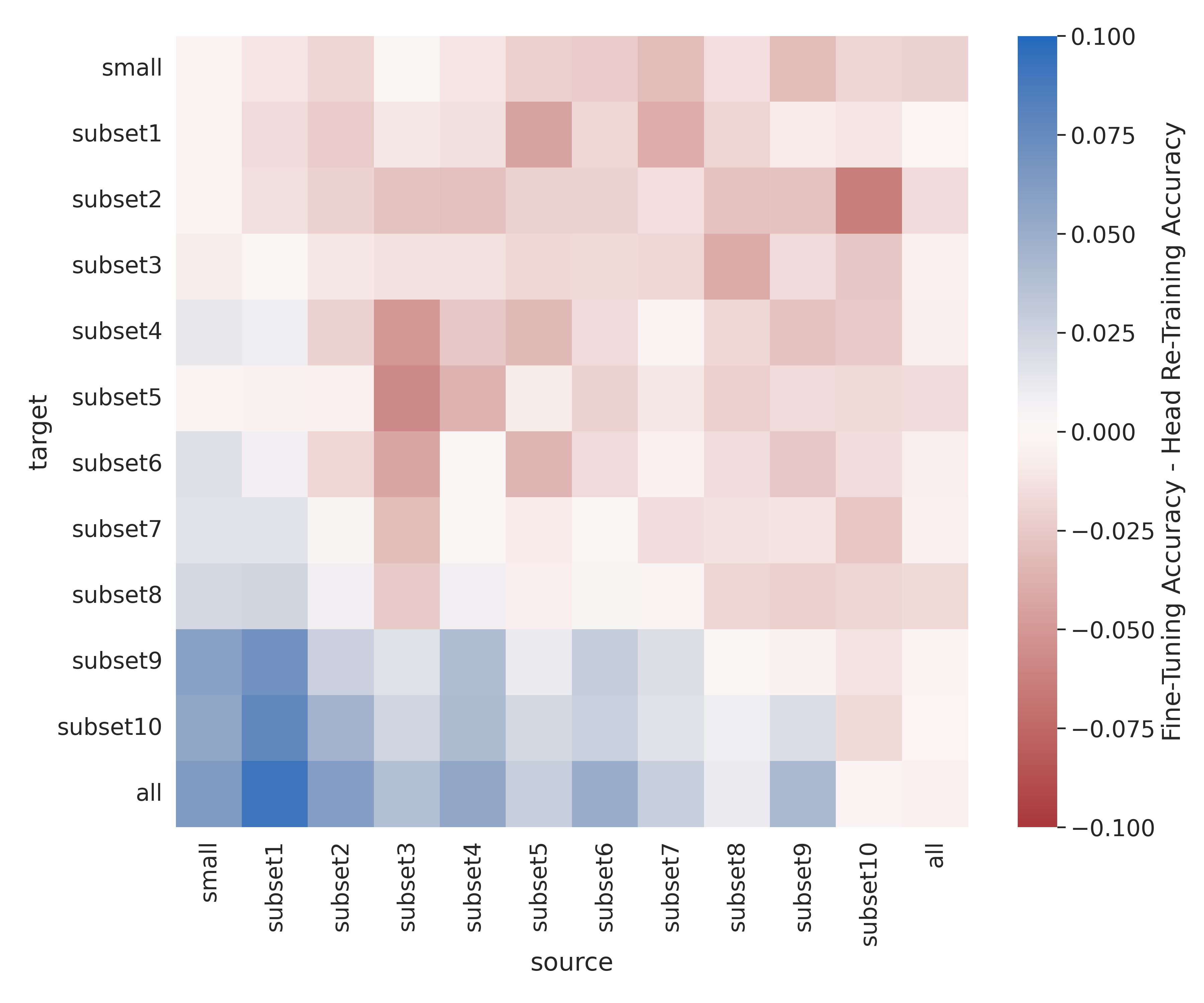}
    \caption{The difference between post-transfer top-1 accuracies achieved using head re-training versus fine-tuning for Modulation Experiment 2, shown on a scale of [-0.1, 0.1]. When the value is positive, fine-tuning outperforms head re-training. When the value is negative, head re-training outperforms fine-tuning.}
    \label{fig:modgroup2_method_diff}
\end{figure}
\unskip

\subsubsection{Head Re-training vs. Fine Tuning}
Finally, Figs. \ref{fig:modgroup1_method_diff} and \ref{fig:modgroup2_method_diff} present the difference between post-transfer top-1 accuracies achieved using head re-training versus fine-tuning for Modulation Experiments 1 and 2 respectively.
These results show that fine-tuning outperforms head re-training in all cases where the source/target tasks are `less similar."
Meanwhile, when the source/target subsets have some modulation schemes in common, head re-training outperforms fine-tuning.
However, as was the case in the domain adaptation experiments discussed in Section \ref{sec:da_method_comparison}, head re-training is as effective, if not more effective, than fine-tuning in the settings where \ac{TL} increases performance over the baseline.
Therefore, head re-training is the preferred method of performing \ac{RF} sequential learning as well, as head re-training is generally more time efficient and less computationally expensive than fine-tuning.

%% file: text/conclusion.tex
\Ac{TL} has yielded tremendous performance benefits in \ac{CV} and \ac{NLP}, and as a result, \ac{TL} is all but commonplace in these fields.
However, the benefits of \ac{TL} have yet to be fully demonstrated and integrated in \ac{RFML}.
To begin to address this deficit, this work systematically evaluated \ac{RF} \ac{TL} performance as a function of \ac{SNR}, \ac{FO}, and modulation type for an \ac{AMC} use-case.
Through this exhaustive study, a number of guidelines have been identified for when and how to use \ac{RF} \ac{TL} successfully.
More specifically, results indicate:
\begin{easylist}[itemize]
@ Using source models trained on the most similar domain/task to the target yields highest performance
@ Transferring from a more challenging domain/task than the target, is preferred to transferring from an easier domain/task
@ Selecting source models based on the similarity of the source/target channel environment is more important than the similarity of the source/target platform(s)
@ Head re-training generally provides the highest performance in any \ac{RF} \ac{TL} case where \ac{TL} provides a performance benefit over training from random initialization, as measured by post-transfer top-1 accuracy, time efficiency, and computational complexity. 
\end{easylist}

As previously mentioned these initial guidelines are subject to further experimentation using additional signal types, channel models, use-cases, model architectures, and augmented or captured datasets.
Continuing and extending the analysis conducted herein will provide a more thorough understanding of \ac{RF} \ac{TL} behavior and performance across a wider range of use-cases and deployment settings.
Further experimentation should include:
\begin{easylist}[itemize]
@ An analysis of multi-task learning behavior using synthetic and/or captured data
@ Analyses of \ac{RF} \ac{TL} performance across other metadata parameters-of-interest such as fading/multi-path channel environments, sample rate, \ac{IQ} imbalance, etc.
@ Analyses of \ac{TL} performance for other \ac{RFML} use-cases such as \ac{SEI}, signal detection, etc.
@ An analysis of \ac{RF} \ac{TL} techniques for transferring between use-cases. For example, can sequential learning techniques be used to transfer between \ac{AMC} and \ac{SEI} use-cases? Is multi-task learning better suited to performing this type of transfer?
@ Analyses of \ac{RF} \ac{TL} performance across varying domains/tasks using captured data
@ An analysis of \ac{RF} \ac{TL} performance across synthetic, augmented, and captured datasets
\end{easylist}

Provided future verification and refinement of these results and guidelines, these guidelines can be used in future \ac{RFML} systems to construct the highest performing models for a given target domain when data is limited.
More specifically, these guidelines begin a discussion regarding how best to continually update \ac{RFML} models once deployed, in an online or incremental fashion, to overcome the highly fluid nature of modern communication systems \cite{wong2021ecosystem}.

%% file: template.bbl
\begin{thebibliography}{999}

\bibitem[{Morocho-Cayamcela} \em{et~al.}(2019){Morocho-Cayamcela}, {Lee}, and
  {Lim}]{5g}
{Morocho-Cayamcela}, M.E.; {Lee}, H.; {Lim}, W.
\newblock Machine Learning for {5G/B5G} Mobile and Wireless Communications:
  Potential, Limitations, and Future Directions.
\newblock {\em IEEE Access} {\bf 2019}, {\em 7},~137184--137206.
\newblock {\url{https://doi.org/10.1109/ACCESS.2019.2942390}}.

\bibitem[Wong \em{et~al.}(2021)Wong, Clark, Flowers, Buehrer, Headley, and
  Michaels]{wong2021ecosystem}
Wong, L.J.; Clark, W.H.; Flowers, B.; Buehrer, R.M.; Headley, W.C.; Michaels,
  A.J.
\newblock An {RFML} Ecosystem: Considerations for the Application of Deep
  Learning to Spectrum Situational Awareness.
\newblock {\em IEEE Open Journal of the Communications Society} {\bf 2021},
  {\em 2},~2243--2264.
\newblock {\url{https://doi.org/10.1109/OJCOMS.2021.3112939}}.

\bibitem[Clark~IV \em{et~al.}(2021)Clark~IV, Hauser, Headley, and
  Michaels]{clark2020}
Clark~IV, W.H.; Hauser, S.; Headley, W.C.; Michaels, A.J.
\newblock Training data augmentation for deep learning radio frequency systems.
\newblock {\em The Journal of Defense Modeling and Simulation} {\bf 2021}, {\em
  18},~217--237.

\bibitem[Hauser(2018)]{hauser2018}
Hauser, S.C.
\newblock Real-World Considerations for Deep Learning in Spectrum Sensing.
\newblock Master's thesis, Virginia Tech,  2018.

\bibitem[Sankhe \em{et~al.}(2019)Sankhe, Belgiovine, Zhou, Riyaz, Ioannidis,
  and Chowdhury]{sankhe2019}
Sankhe, K.; Belgiovine, M.; Zhou, F.; Riyaz, S.; Ioannidis, S.; Chowdhury, K.
\newblock {ORACLE: Optimized Radio Classification through Convolutional Neural
  Networks}.
\newblock In Proceedings of the IEEE INFOCOM 2019-IEEE Conference on Computer
  Communications. IEEE,  2019, pp. 370--378.

\bibitem[Ruder(2019)]{ruder2019}
Ruder, S.
\newblock Neural transfer learning for natural language processing.
\newblock PhD thesis, NUI Galway,  2019.

\bibitem[Olivas \em{et~al.}(2009)Olivas, Guerrero, Martinez-Sober,
  Magdalena-Benedito, Serrano, et~al.]{olivas2009}
Olivas, E.S.; Guerrero, J.D.M.; Martinez-Sober, M.; Magdalena-Benedito, J.R.;
  Serrano, L.;  et~al.
\newblock {\em Handbook of research on machine learning applications and
  trends: Algorithms, methods, and techniques}; IGI Global,  2009.

\bibitem[Wong and Michaels(2022)]{wong2022}
Wong, L.J.; Michaels, A.J.
\newblock Transfer Learning for Radio Frequency Machine Learning: A Taxonomy
  and Survey.
\newblock {\em Sensors} {\bf 2022}, {\em 22}.
\newblock {\url{https://doi.org/10.3390/s22041416}}.

\bibitem[Rondeau()]{rfmls}
Rondeau, T.
\newblock {Radio Frequency Machine Learning Systems (RFMLS)}.

\bibitem[{Dobre} \em{et~al.}(2007){Dobre}, {Abdi}, {Bar-Ness}, and
  {Su}]{dobre2007}
{Dobre}, O.A.; {Abdi}, A.; {Bar-Ness}, Y.; {Su}, W.
\newblock Survey of automatic modulation classification techniques: classical
  approaches and new trends.
\newblock {\em IET Communications} {\bf 2007}, {\em 1},~137--156.

\bibitem[{West} and {O'Shea}(2017)]{west2017}
{West}, N.E.; {O'Shea}, T.
\newblock Deep architectures for modulation recognition.
\newblock In Proceedings of the 2017 IEEE Int. Symp. on Dynamic Spectrum Access
  Networks (DySPAN),  2017, pp. 1--6.

\bibitem[{Pan} and {Yang}(2010)]{pan2010}
{Pan}, S.J.; {Yang}, Q.
\newblock A Survey on Transfer Learning.
\newblock {\em IEEE Trans. on Knowledge and Data Eng.} {\bf 2010}, {\em
  22},~1345--1359.
\newblock {\url{https://doi.org/10.1109/TKDE.2009.191}}.

\bibitem[{Chen} \em{et~al.}(2019){Chen}, {Zheng}, {Yang}, and {Yang}]{chen2019}
{Chen}, S.; {Zheng}, S.; {Yang}, L.; {Yang}, X.
\newblock Deep Learning for Large-Scale Real-World {ACARS} and {ADS-B} Radio
  Signal Classification.
\newblock {\em IEEE Access} {\bf 2019}, {\em 7},~89256--89264.
\newblock {\url{https://doi.org/10.1109/ACCESS.2019.2925569}}.

\bibitem[{Pati} \em{et~al.}(2020){Pati}, {Kaneko}, and
  {Taparugssanagorn}]{pati2020}
{Pati}, B.M.; {Kaneko}, M.; {Taparugssanagorn}, A.
\newblock A Deep Convolutional Neural Network Based Transfer Learning Method
  for Non-Cooperative Spectrum Sensing.
\newblock {\em IEEE Access} {\bf 2020}, {\em 8},~164529--164545.
\newblock {\url{https://doi.org/10.1109/ACCESS.2020.3022513}}.

\bibitem[Kuzdeba \em{et~al.}(2021)Kuzdeba, Robinson, and Carmack]{kuzdeba2021}
Kuzdeba, S.; Robinson, J.; Carmack, J.
\newblock Transfer Learning with Radio Frequency Signals.
\newblock In Proceedings of the 2021 IEEE 18th Annual Consumer Communications
  Networking Conference (CCNC),  2021, pp. 1--9.
\newblock {\url{https://doi.org/10.1109/CCNC49032.2021.9369550}}.

\bibitem[Robinson and Kuzdeba(2021)]{robinson2021}
Robinson, J.; Kuzdeba, S.
\newblock {RiftNet}: Radio Frequency Classification for Large Populations.
\newblock In Proceedings of the 2021 IEEE 18th Annual Consumer Communications
  Networking Conference (CCNC),  2021, pp. 1--6.
\newblock {\url{https://doi.org/10.1109/CCNC49032.2021.9369455}}.

\bibitem[{O’Shea} \em{et~al.}(2018){O’Shea}, {Roy}, and
  {Clancy}]{oshea2018}
{O’Shea}, T.J.; {Roy}, T.; {Clancy}, T.C.
\newblock Over-the-Air Deep Learning Based Radio Signal Classification.
\newblock {\em IEEE Journal of Selected Topics in Signal Processing} {\bf
  2018}, {\em 12},~168--179.
\newblock {\url{https://doi.org/10.1109/JSTSP.2018.2797022}}.

\bibitem[{Dörner} \em{et~al.}(2018){Dörner}, {Cammerer}, {Hoydis}, and
  t.~{Brink}]{dorner2018}
{Dörner}, S.; {Cammerer}, S.; {Hoydis}, J.; t.~{Brink}, S.
\newblock Deep Learning Based Communication Over the Air.
\newblock {\em IEEE Journal of Selected Topics in Signal Processing} {\bf
  2018}, {\em 12},~132--143.
\newblock {\url{https://doi.org/10.1109/JSTSP.2017.2784180}}.

\bibitem[{Zheng} \em{et~al.}(2020){Zheng}, {Chen}, {Qi}, {Zhou}, and
  {Yang}]{zheng2020}
{Zheng}, S.; {Chen}, S.; {Qi}, P.; {Zhou}, H.; {Yang}, X.
\newblock Spectrum sensing based on deep learning classification for cognitive
  radios.
\newblock {\em China Comm.} {\bf 2020}, {\em 17},~138--148.
\newblock {\url{https://doi.org/10.23919/JCC.2020.02.012}}.

\bibitem[Clark \em{et~al.}(2019)Clark, Leffke, Headley, and Michaels]{cyborg}
Clark, B.; Leffke, Z.; Headley, C.; Michaels, A.
\newblock Cyborg Phase {II} Final Report.
\newblock Technical report, Ted and Karyn Hume Center for National Security and
  Technology,  2019.

\bibitem[{Peng} \em{et~al.}(2020){Peng}, {Gilman}, {Vasconcelos}, {Cosman}, and
  {Milstein}]{peng2020}
{Peng}, Q.; {Gilman}, A.; {Vasconcelos}, N.; {Cosman}, P.C.; {Milstein}, L.B.
\newblock Robust Deep Sensing Through Transfer Learning in Cognitive Radio.
\newblock {\em IEEE Wireless Comm. Letters} {\bf 2020}, {\em 9},~38--41.
\newblock {\url{https://doi.org/10.1109/LWC.2019.2940579}}.

\bibitem[{Ye} \em{et~al.}(2020){Ye}, {Li}, {Yu}, {Zhao}, {Liu}, and
  {Hou}]{ye2020}
{Ye}, N.; {Li}, X.; {Yu}, H.; {Zhao}, L.; {Liu}, W.; {Hou}, X.
\newblock Deep{NOMA}: A Unified Framework for {NOMA} Using Deep Multi-Task
  Learning.
\newblock {\em IEEE Trans. on Wireless Comm.} {\bf 2020}, {\em 19},~2208--2225.
\newblock {\url{https://doi.org/10.1109/TWC.2019.2963185}}.

\bibitem[{Clark} \em{et~al.}(2019){Clark}, {Arndorfer}, {Tamir}, {Kim},
  {Vives}, {Morris}, {Wong}, and {Headley}]{clark2019}
{Clark}, W.H.; {Arndorfer}, V.; {Tamir}, B.; {Kim}, D.; {Vives}, C.; {Morris},
  H.; {Wong}, L.; {Headley}, W.C.
\newblock Developing {RFML} Intuition: An Automatic Modulation Classification
  Architecture Case Study.
\newblock In Proceedings of the 2019 IEEE Military Comm. Conference (MILCOM),
  2019, pp. 292--298.
\newblock {\url{https://doi.org/10.1109/MILCOM47813.2019.9020949}}.

\bibitem[Wong and McPherson(2021)]{wong2021}
Wong, L.J.; McPherson, S.
\newblock Explainable Neural Network-based Modulation Classification via
  Concept Bottleneck Models.
\newblock In Proceedings of the 2021 IEEE Computing and Comm. Workshop and
  Conference (CCWC),  2021.

\bibitem[{Clark IV} \em{et~al.}(2021){Clark IV}, Hauser, Headley, and
  Michaels]{clark2021}
{Clark IV}, W.H.; Hauser, S.; Headley, W.C.; Michaels, A.J.
\newblock Training Data Augmentation for Deep Learning Radio Frequency Systems.
\newblock {\em {JDMS} Special Issue} {\bf 2021}.

\bibitem[Merchant(2019)]{merchant2019}
Merchant, K.
\newblock Deep Neural Networks for Radio Frequency Fingerprinting.
\newblock PhD thesis,  2019.

\bibitem[Liu \em{et~al.}(2020)Liu, Lian, Farrell, and Wandell]{liu2020}
Liu, Z.; Lian, T.; Farrell, J.; Wandell, B.A.
\newblock Neural network generalization: The impact of camera parameters.
\newblock {\em IEEE Access} {\bf 2020}, {\em 8},~10443--10454.

\bibitem[Gaeddert()]{liquid}
Gaeddert, J.
\newblock liquid-dsp.

\bibitem[Wong \em{et~al.}(2022)Wong, McPherson, and
  Michaels]{wong_dataset_2022}
Wong, L.J.; McPherson, S.; Michaels, A.J.
\newblock {Transfer Learning for RF Domain Adaptation – Synthetic Dataset},
  2022.

\bibitem[Hilburn \em{et~al.}(2018)Hilburn, West, O'Shea, and Roy]{hilburn2018}
Hilburn, B.; West, N.; O'Shea, T.; Roy, T.
\newblock {SigMF}: the signal metadata format.
\newblock In Proceedings of the Proceedings of the GNU Radio Conference,  2018,
  Vol.~3.

\bibitem[Wong and McPherson(2021)]{wong2021cb}
Wong, L.J.; McPherson, S.
\newblock Explainable Neural Network-based Modulation Classification via
  Concept Bottleneck Models.
\newblock In Proceedings of the 2021 IEEE 11th Annual Computing and
  Communication Workshop and Conference (CCWC),  2021, pp. 0191--0196.
\newblock {\url{https://doi.org/10.1109/CCWC51732.2021.9376108}}.

\bibitem[O’Shea \em{et~al.}(2016)O’Shea, Corgan, and Clancy]{oshea2016}
O’Shea, T.J.; Corgan, J.; Clancy, T.C.
\newblock Convolutional radio modulation recognition networks.
\newblock In Proceedings of the International conference on engineering
  applications of neural networks. Springer,  2016, pp. 213--226.

\bibitem[Kingma and Ba(2014)]{kingma2014}
Kingma, D.P.; Ba, J.
\newblock Adam: A method for stochastic optimization.
\newblock {\em arXiv preprint arXiv:1412.6980} {\bf 2014}.

\bibitem[pyt()]{pytorchCE}
Cross Entropy Loss.

\bibitem[Paszke \em{et~al.}(2019)Paszke, Gross, Massa, Lerer, Bradbury, Chanan,
  Killeen, Lin, Gimelshein, Antiga, et~al.]{pytorch}
Paszke, A.; Gross, S.; Massa, F.; Lerer, A.; Bradbury, J.; Chanan, G.; Killeen,
  T.; Lin, Z.; Gimelshein, N.; Antiga, L.;  et~al.
\newblock {PyTorch}: An imperative style, high-performance deep learning
  library.
\newblock {\em Advances in neural information processing systems} {\bf 2019},
  {\em 32},~8026--8037.

\bibitem[Rosenstein \em{et~al.}(2005)Rosenstein, Marx, Kaelbling, and
  Dietterich]{rosenstein2005}
Rosenstein, M.T.; Marx, Z.; Kaelbling, L.P.; Dietterich, T.G.
\newblock To transfer or not to transfer.
\newblock In Proceedings of the NIPS 2005 workshop on transfer learning,  2005,
  Vol. 898, pp. 1--4.

\end{thebibliography}
